\documentclass[journal=mamobx,manuscript=article]{achemso}

\usepackage[version=3]{mhchem} 
\usepackage[T1]{fontenc}       

\usepackage{graphicx}
\usepackage{dcolumn}
\usepackage{bm}
\usepackage{amsmath}
\usepackage{amsfonts}
\usepackage{amssymb}
\usepackage{braket}
\usepackage[separate-uncertainty = true,multi-part-units=single]{siunitx}
\usepackage{subcaption}
\usepackage{color}

\usepackage{newfloat}
\DeclareFloatingEnvironment[name={Supporting Figure}]{suppfigure}

\let\oldsqrt\sqrt
\def\sqrt{\mathpalette\DHLhksqrt}
\def\DHLhksqrt#1#2{%
\setbox0=\hbox{$#1\oldsqrt{#2\,}$}\dimen0=\ht0
\advance\dimen0-0.2\ht0
\setbox2=\hbox{\vrule height\ht0 depth -\dimen0}%
{\box0\lower0.4pt\box2}}

\title{Polymer brush collapse under shear flow}

\author{Airidas Korolkovas}
\affiliation{Institut Laue-Langevin, 71 rue des Martyrs, 38000 Grenoble, France}
\alsoaffiliation{Universit\'{e} Grenoble Alpes, Liphy, 140 Rue de la Physique, 38402 Saint-Martin-d'H\`{e}res, France}
\email{korolkovas@ill.fr}

\author{Cesar Rodriguez-Emmenegger}
\affiliation{DWI - Leibniz Institute for Interactive Materials and Institute of Technical and Macromolecular Chemistry, RWTH Aachen University, Forckenbeckstra{\ss}e 50, 52074 Aachen, Germany}

\author{Andres de los Santos Pereira}
\affiliation{Institute of Macromolecular Chemistry, Academy of Sciences of the Czech Republic v.v.i., Heyrovsky Sq. 2, 162 06 Prague, Czech Republic}

\author{Alexis Chennevi\`ere}
\affiliation{Laboratoire L\'eon Brillouin, CEA, CNRS, Universit\'e Paris-Saclay, Saclay 91191 Gif-sur-Yvette Cedex, France.}

\author{Fr\'ed\'eric Restagno}
\affiliation{Laboratoire de Physique des Solides, CNRS, Univ. Paris-Sud, Universit\'e Paris-Saclay, 91405 Orsay Cedex, France}

\author{Maximilian Wolff}
\affiliation{Division for Material Physics, Department for Physics and Astronomy, Uppsala University, Box 516, 75120 Uppsala, Sweden}

\author{Franz A. Adlmann}
\affiliation{Division for Material Physics, Department for Physics and Astronomy, Uppsala University, Box 516, 75120 Uppsala, Sweden}

\author{Andrew J.C. Dennison}
\affiliation{Department of Chemistry, Technical University Berlin, 10623 Berlin, Germany}
\alsoaffiliation{Institut Laue-Langevin, 71 rue des Martyrs, 38000 Grenoble, France}
\alsoaffiliation{Department of Physics and Astronomy, University of Sheffield, S102TN Sheffield, UK}

\author{Philipp Gutfreund}
\affiliation{Institut Laue-Langevin, 71 rue des Martyrs, 38000 Grenoble, France}
\email{gutfreund@ill.fr}

\begin{document}




\begin{abstract}
Shear responsive surfaces offer potential advances in a number of applications. Surface functionalisation using polymer brushes is one route to such properties, particularly in the case of entangled polymers. We report on neutron reflectometry measurements of polymer brushes in entangled polymer solutions performed under controlled shear, as well as coarse-grained computer simulations corresponding to these interfaces. Here we show a reversible and reproducible collapse of the brushes, increasing with the shear rate. Using two brushes of greatly different chain lengths and grafting densities, we demonstrate that the dynamics responsible for the structural change of the brush are governed by the free chains in solution rather than the brush itself, within the range of parameters examined. The phenomenon of the brush collapse could find applications in the tailoring of nanosensors, and as a way to dynamically control surface friction and adhesion.
\end{abstract}

\maketitle

\section{Introduction}
A polymer brush is a unique type of surface functionalisation, consisting of long polymer chains densely tethered by one end to a surface~\cite{Milner1991, Brittain_2007}. The conformation of a solvated polymer brush is markedly different to that of chains in bulk polymer solution as the brush must stretch away from the surface to minimize contact with the densely grafted neighbouring chains. Polymer brushes have broad interest across a variety of sectors since tuning interfacial properties (e.g. chemical composition, molecular weight, grafting density) can yield surface coatings with a high degree of control and in some cases completely new functionality.

One of the most common uses for brushes is to inhibit protein adsorption and prevent surface fouling~\cite{schottler2016protein, butcher2016drug}. Various other applications are also under investigation~\cite{Azzaroni2012} ranging from bioactive interfaces~\cite{Bauer2015}, to brush-mediated lubrication~\cite{raviv2003lubrication,Klein2009}, to soil release in textiles~\cite{Yang_2012}, and even semiconductor manufacturing~\cite{Pinto_2008, Youm_2016}. Another emerging application is the use of polymer brushes as nanosensors reacting to various stimuli including pressure~\cite{Reinhardt2013}, light~\cite{Brown_2009}, temperature~\cite{Lutz_2011}, and pH~\cite{tokareva2004nanosensors}, among others. Remarkably, the sensitivity of these nanoscale sensors can be finely tuned by the amount of the brush swelling~\cite{Stuart2010}, which in turn depends on the nature of the solvent, but can also be affected by other factors such as shear stress as will be shown in this article.

\begin{figure}[!tbh]
\begin{minipage}{\textwidth}
	\centering
		\begingroup
			\sbox0{\includegraphics{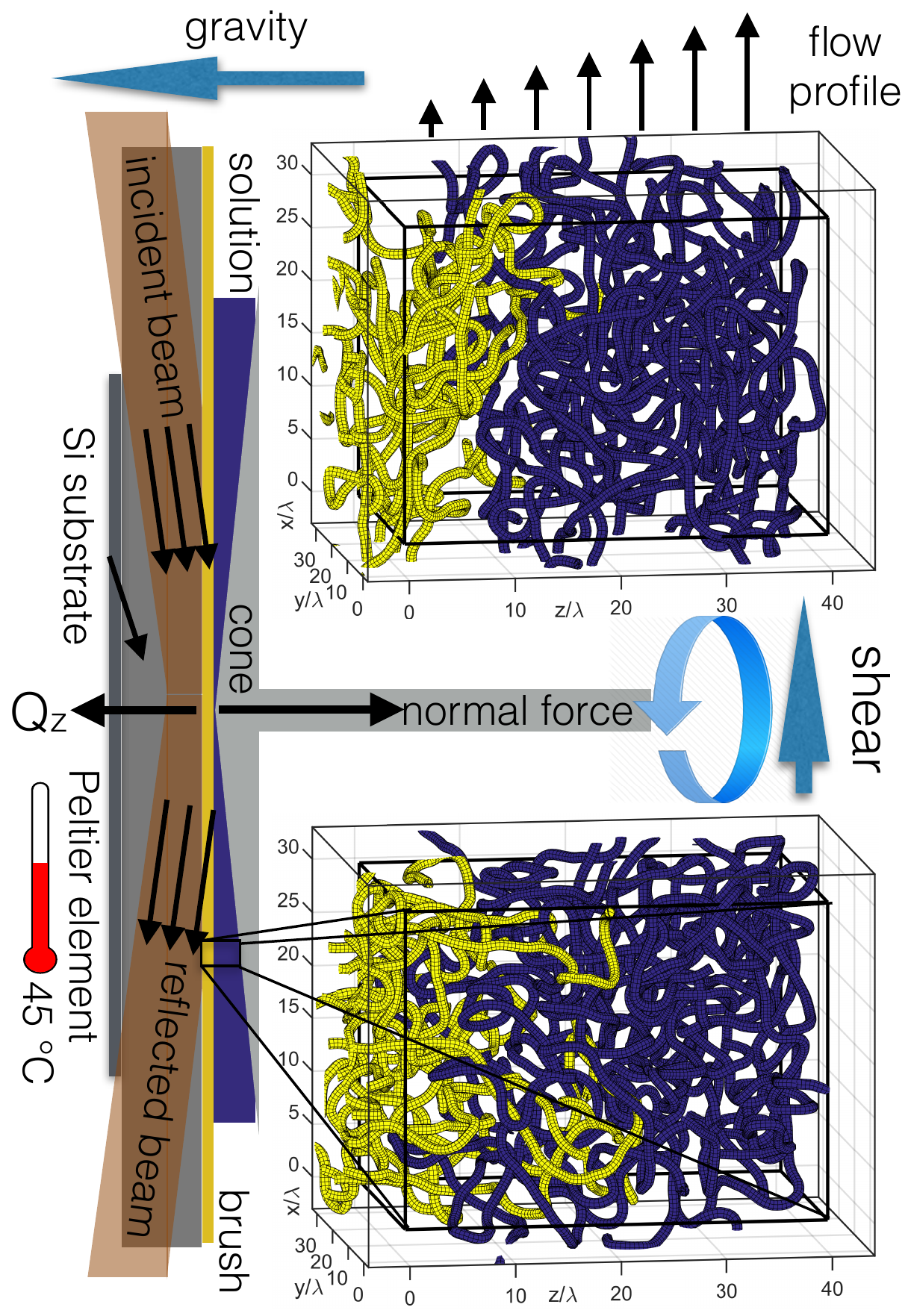}} 
			\includegraphics[clip,trim={.0\wd0} 0 {.0\wd0} 0,width=0.5\linewidth]{./fignew/geometry.pdf} 
		\endgroup
    \caption{Experimental setup and simulated polymer conformation. The experiments are conducted with the silicon-polymer interface horizontally oriented. The wavevector transfer $Q_z$ is perpendicular to the interface. The temperature of the silicon substrate is controlled by a Peltier element from the bottom side. Shear is applied by the rheometer \emph{via} a titanium cone or plate. The lower right panel shows a conformation snapshot plotted from the simulation data. The interpenetration of the polymer brush (yellow) and free chains (blue) is clearly visible. With applied shear (upper right panel) the free chains are pulled out of the brush, the mean thickness of the brush decreases and the interface becomes sharper. The density profiles along the $z$-direction are compared to the neutron reflectometry data.}\label{static3D}
\end{minipage}
\end{figure}

The static properties of polymer brushes are well understood thanks to extensive theoretical~\cite{milner1988theory}, computer simulation~\cite{murat1989interaction, binder2012polymer}, and experimental~\cite{kelley1998direct,Jones1999,Currie2003,Kato2003} studies. The knowledge of brush dynamics, however, is still incomplete, even though it is crucial for the design of the aforementioned sensors. To help fill this demand, our study will focus on the response of brushes to an applied shear stress while swollen and deeply interpenetrated with a bulk polymer solution, illustrated in Fig~\ref{static3D}. Aside from use in sensors, surfaces decorated with brushes may also play a key role to control adhesion~\cite{malham2009density, la2007controlling}, lubrication~\cite{Durliat1997}, friction~\cite{Leger1999, Cohen2011}, and in microfluidic devices~\cite{Thorsen2002} and confined channels~\cite{Raviv2003}. In our experimental conditions, the chains are strongly entangled reaching a relaxation time on the order of $\tau_d = \SI{1}{\second}$, which has immediate practical importance since it can dynamically interact with the flows encountered in the aforementioned real world applications, which commonly have similar time scales. However, the brush-bulk interface remains very challenging to investigate either theoretically~\cite{larson2015modeling} or experimentally, due to its complex, heterogeneous, strongly interacting, non-equilibrium, and confined nature. We have therefore taken a two pronged approach and used a recently developed computer simulation technique~\cite{korolkovas2016simulation} as well as state-of-the-art experimental rheology - neutron reflectometry (rheo-NR)~\cite{Wolff2013} capabilities. This combination enables greater insight into what is occurring at the interface compared to the two approaches taken separately.

Simulation of polymer brushes under shear~\cite{grest1999normal, binder2011polymer} is a vibrant field: brushes in good solvent~\cite{muller2007cyclic, singh2015polymer}, two opposing polyzwitterionic brushes~\cite{mendoncca2016monte}, brushes in contact with short melt chains~\cite{pastorino2006static}, and stiff brushes related to biological membranes~\cite{romer2015dense}, just to name a few recent publications. However, most of the simulations (molecular dynamics, dissipative particle dynamics, and various kinds of Monte Carlo) are based on $\lambda \approx \SI{1}{\nano\meter}$ size beads running at time steps of about $\tau_m = \SI{e-12}{\second}$, required to follow the thermal fluctuations of the bead momentum. Current computers can typically perform $10^8$ time steps within a reasonable execution time; insufficient to bridge the gap to our experimental goal of $\tau_d = \SI{1}{\second}$.

The next level of coarse-graining is the Brownian dynamics where we abandon the bead momentum altogether and only track their positions, which take about $\tau = \SI{e-9}{\second}$ to relax after diffusing a distance greater than their own size. This technique has already been used to predict a brush collapse under shear~\cite{saphiannikova1998self}.  However, the reported collapse occurred at a shear rate approaching $\dot{\gamma} \approx 1/\tau \approx \SI{e9}{\per\second}$ and was due to the finite extensibility of the polymer backbone. Such extreme shear rates are more akin to an explosion than a well controlled shear experiment and this mode of brush collapse is not related to the entanglement dynamics at $\dot{\gamma} \approx 1/\tau_d \approx \SI{1}{\per\second}$ relevant to realistic flow conditions measured in our study.

Our experiments are done using polystyrene (PS) in a good solvent at $\phi=\SI{30}{\percent}$ fraction by weight. To describe this liquid, the appropriate coarse particle is called a blob~\cite{deGennesScalingConcepts} and its size corresponds to the typical distance between neighbouring polymer chains: $\lambda = a\phi^{-3/4}$, where $a \approx \SI{7}{\angstrom}$ is the size of one styrene monomer. The blob repulsion is best quantified by an effective Gaussian potential which results in the correct static structure~\cite{bolhuis2001accurate}. Dynamically, however, this blob potential was considered too weak and too soft to prevent chain crossings~\cite{nikunen2007reptational} and therefore unable to produce any entanglements. A recent study~\cite{korolkovas2016simulation}, however, has proposed to smear out the Gaussian potential in both time and space, thus suppressing chain crossings while retaining the long Brownian time step $\tau$ adequate to describe our experiments.

Neutron Reflectometry (NR) is a powerful experimental tool for the structural and dynamical investigation of polymer brushes, thanks to the possibility of isotopic replacement to enhance the contrast between the grafted and the bulk polymers, as well as its atomic resolution and non-invasive nature. A unique advantage of NR is that most engineering materials like aluminium or silicon are transparent for the neutrons which permits direct measurement of the brush-bulk interface through the silicon substrate~\cite{Wolff2013}. Structural investigations of brushes under shear load have been performed by NR measurements on PS brushes in solvents~\cite{Baker2000,Ivkov2001}, but found no measurable effect. Next, we look at two studies which examined a PS brush in contact with a PS melt. The first one was measured \textit{in situ} while shearing~\cite{Sasa2011}. No reproducible result could be obtained and it was explained by metastable states of the brush. However, very high torques were applied in that study and the brushes were not characterised after the shear experiments. It has been shown by NR that PS brushes can be destroyed by high torque shear~\cite{WolffGutfreund2011} and such a scenario is likely in the aforementioned experiment. The second study also sheared PS brushes in a PS melt~\cite{Chenneviere2016}, which were then rapidly quenched below their glass transition temperature and measured \emph{ex situ} with NR, reporting a reproducible retraction of the brush. In our present study we have used NR for an \emph{in situ} characterisation of the behaviour of PS brushes under shear by an entangled PS solution in diethyl phthalate (DEP, a good solvent of very low volatility). The use of solution rather than melt is more relevant to biological processes as well as microfluidic applications.

Here we show both experimentally and computationally that the entangled polymer brush thickness decreases with shear. More precisely, we observe a shrinking of brushes proportional to the square of the applied shear rate. This non-linear effect is attributed to the normal stress difference, which is an excess pressure buildup perpendicular to the applied shear flow, and is well-known to occur in bulk entangled polymer fluids, where it leads to the so-called Weissenberg effect~\cite{Weissenberg1947}. The time scale of the brush collapse is determined by the reptation time of the free chains in solution, rather than the internal dynamics of the brush. The brush thickness returns to equilibrium upon cessation of shear, and the effect can by cycled many times over. The experimental and simulation findings are in good agreement and are further corroborated by a simple phenomenological theory.

\section{Experimental}
\subsection{Materials}
\textit{N,N,N',N'',N''}-Pentamethyldiethylenetriamine (PMDETA, \SI{99}{\percent}), styrene (\SI{99}{\percent}), diethoxy(3-glycidyloxypropyl)methysilane (\SI{99}{\percent}), dichloromethane (\SI{99}{\percent}) and diethyl phthalate (DEP) (\SI{99}{\percent}) were purchased from Sigma-Aldrich (Czech Republic). Deuterated polystyrene (dPS), $M_w = \SI{627}{\kilogram\per\mole}$, $M_w/M_n = 1.09$, correspopnding to $P = M_w/\SI{112.2}{\gram\per\mole} = 5570$, was purchased from Polymer Source, Canada. Monocrystalline silicon blocks of size $7\times7\times1\, \text{cm}$, orientation $(1,\,0,\,0)$, were purchased from CrysTec, Germany. Styrene was distilled over CaH${}_2$ under reduced pressure and stored under Ar.\newline
[11-(2-Bromo-2-methyl)propionyloxy]undecyltrichlorosilane was synthesized according to a previously published protocol~\cite{rodriguez2015quantifying}.

\subsection{Preparation of Brush-long-sparse: ``grafting-to'' approach}
The amino end-functionalized PS was synthesized in-house to a molecular weight of $M_{n}=218$\,kg/mol ($N = M_n/\SI{104.15}{\gram\per\mole} = 2093$) and a polydispersity of 1.23. Then it was grafted onto a self-assembled monolayer (SAM) of diethoxy(3-glycidyloxypropyl)methysilane)  deposited on a single crystal silicon block. Details about the sample preparation can be found in Ref.~ \cite{Chenneviere2013}. The thickness of the SAMs was determined by ellipsometry and found to be \SI{1.0}{\nano\meter} for both brushes corresponding to fully stretched and upright standing chains in accord with previous samples~\cite{Chenneviere2013}. The silicon oxide thickness was determined by NR as described in the SI.

\subsection{Preparation of Brush-short-dense: ``grafting-from'' approach}
PS brushes were grafted from an initiator-coated substrate by surface-initiated atom transfer radical polymerization (ATRP) employing a literature procedure~\cite{ell2009structural}, modified to achieve a lower grafting density and high thickness. Firstly, a self-assembled monolayer of ATRP initiator was immobilized on the surface. The substrate (silicon slab) was rinsed with toluene, acetone, ethanol, and deionized water, blown dry with nitrogen, and activated in a UV/O${}_3$ cleaner for \SI{20}{\minute}. Without delay, the sample was placed in a custom-made reactor vessel, which was then sealed, evacuated, and refilled with Ar. A \SI{1}{\micro\gram\per\milli\liter} solution of (11-(2-bromo-2-methyl)propionyloxy)undecyltrichlorosilane in anhydrous toluene was added until the sample was fully immersed. The immobilization of the initiator was allowed to proceed for \SI{3}{\hour} at room temperature and the sample was subsequently removed from the reactor, rinsed copiously with toluene, acetone, ethanol, and deionized water, and dried by blowing with nitrogen.

To achieve a lowered grafting density, a fraction of the surface-grafted ATRP initiator groups were deactivated by nucleophilic substitution with NaN${}_3$. The sample was placed in a custom-made reactor, which was then sealed, evacuated, and refilled with Ar, and placed in a thermostatic bath at \SI{60}{\celsius} for \SI{1}{\hour} to reach thermal equilibrium. A solution of NaN${}_3$ (\SI{3.4}{\milli\gram\per\milli\liter}) in anhydrous \textit{N,N}-dimethylformamide (DMF), previously heated to \SI{60}{\celsius}, was added to completely cover the sample and the reaction was allowed to proceed at \SI{60}{\celsius} for \SI{8}{\hour}. Subsequently, the reaction was stopped by replacing the solution in the reactor with pure DMF. The sample was removed from the reactor, rinsed copiously with DMF, ethanol, and deionized water, and dried by carefully blowing with nitrogen.

For the surface-initiated ATRP, styrene (\SI{40}{\milli\liter}, \SI{349}{\milli\mole}), anhydrous toluene (\SI{20}{\milli\liter}), and PMDETA (\SI{760}{\micro\liter}, \SI{3.64}{\milli\mole}) were degassed in Schlenk flask via three freeze-pump-thaw cycles. The solution was transferred under Ar to another Schlenk flask containing CuBr (\SI{496}{\milli\gram}, \SI{3.46}{\milli\mole}) and CuBr${}_2$ (\SI{40}{\milli\gram}, \SI{0.179}{\milli\mole}), which had been previously deoxygenated by three vacuum/Ar-backfilling cycles. The flask containing the polymerization solution was placed in thermostatic bath at \SI{90}{\celsius} and stirred vigorously for \SI{1}{\hour}. The initiator-functionalized substrate was placed vertically in a custom-made reactor, which was subsequently closed, deoxygenated by three cycles of vacuum/Ar-backfilling, and placed in a thermostatic oil bath at \SI{90}{\celsius} to allow the temperature to equilibrate. The polymerization solution was transferred under Ar to the reactor containing the substrate and the reaction was allowed to proceed at \SI{90}{\celsius} for \SI{22}{\hour}. The reaction was stopped by opening the reactor and adding toluene and the substrate was rinsed copiously with toluene, acetone, ethanol, and deionized water and dried by blowing with nitrogen. The dry thickness of the layers was measured by spectroscopic ellipsometry and NR.

\subsection{Rheology}
Deuterated polystyrene (dPS, \SI{0.3}{\gram}) was mixed at \SI{30}{\percent} weight fraction with diethyl phthalate (DEP, \SI{0.7}{\gram}, a good solvent of low volatility), in a round bottom flask. It was topped with an abundant amount (\SI{50}{\milli\liter}) of dichloromethane (also a good solvent, but high volatility), and stirred for several hours to fully dissolve the dPS. The dichloromethane was then slowly removed in a rotary evaporator under reduced pressure, which ensured that no gas bubbles were left trapped in the resulting viscous liquid.

A teflon spatula was used to transfer the dPS-DEP solution onto the brush-coated silicon crystal. The liquid was then contained in an Anton-Paar MCR 501 rheometer in cone-plate or plate-plate geometry (1$^{\circ}$ cone angle, \SI{50}{\milli\meter} diameter for cone or plate) to allow \emph{in situ} rheology as explained in Ref.~\cite{Wolff2013}. The rotating cone or plate on top was made of titanium and its surface was sand-blasted to reduce surface slip at the moving interface. The temperature on the stationary brush-coated side was kept constant at \SI{45}{\celsius} throughout the experiment.

\subsection{Neutron experiment details}
Neutron reflectometry was carried out on FIGARO at the Institut Laue-Langevin, Grenoble, France~\cite{Campbell2011}. The measurements were performed in time-of-flight mode using a wavelength band from 2.2~-~21~\AA\, and a wavelength resolution of \SI{7}{\percent}. Two reflection angles (\SI{0.62}{\degree} and \SI{2.72}{\degree}) were used to cover the full $Q$-range by rotating the incident beam and the detector around the sample keeping the rheometer horizontal at all times. The relative angular divergence was set to $\Delta\theta/\theta = \SI{1.5}{\percent}$ for both reflection angles. The acquisition time was 1~-~5~min for the first reflection angle and 25~min for the second angle and all measurements under shear were reproduced and cycled several times to exclude any transient phenomena. The footprint of the neutron beam (39$\times$35~mm$^{2}$) was centered to the cone/plate, hence the scattering momentum transfer is parallel to the shear gradient. The rheo-NR setup with the neutrons entering through the side of the stationary silicon substrate (see Fig.~\ref{static3D}) is explained in more detail in Ref.~\cite{Wolff2013}.

\section{Simulation method}

Each chain is described by a continuous path $\mathbf{R}(s)$ where $s\in(0,1)$ is the monomer label. The chains have $N$ degrees of freedom and repel one another via a Gaussian potential $\Phi(\mathbf{r}) = k_B T e^{-\mathbf{r}^2/(2\lambda^2)}$, while the backbone stays connected via a harmonic spring interaction of the same strength $k_B T$ and the same length $\lambda$. The continuous backbone $s$ is sampled by a number of $J=4N$ discrete points:
\begin{equation}
\mathbf{R}_j = \mathbf{a}_0 + 2\sum_{n=1}^{N-1}\mathbf{a}_n \cos \left(\frac{\pi(2j-1)n}{2J}\right)
\end{equation}
which ensures that neighbouring points $|\mathbf{R}_j - \mathbf{R}_{j+1}| \ll \lambda$ are closer together than the potential range of the blob $\lambda$, and hence there are effectively no gaps through which the chains could cross. The propagation in time is carried out in terms of $N$ Rouse modes:
\begin{equation}
\mathbf{a}_n(t+\Delta t) = \mathbf{a}_n(t) + (\mathbf{F}_{\text{spring}} + \mathbf{F}_{\text{exvol}})\Delta t + \lambda \sqrt{6\Delta t/(\tau M)}\mathcal{R}_n,
\end{equation}
where standard formulas are used to evaluate the spring and the excluded volume forces. The Brownian time unit can be estimated by the Einstein-Stokes formula:
\begin{equation}
\tau = \frac{6\pi \eta_s \lambda^3}{k_B T} \approx \SI{e-9}{\second},
\end{equation}
where $\eta_s = \SI{1.7e-2}{\pascal\second}$ is the viscosity of DEP.

The important novelty in this simulation is that its time resolution is deliberately truncated by updating the random vector $\mathcal{R}_n$ only at intervals of $M=120$ steps instead of every single $M=1$ step. This ensures that the random force strength is much weaker  than the excluded volume one (by a factor of $\sqrt{M}$), thereby suppressing any chances of chain crossings and giving rise to entanglement dynamics.

Here we note that the maximum applicable shear rate is also limited to about $\dot{\gamma}(M\tau) \ll 1$, and the fastest one we have used was $\text{Wi} = \dot{\gamma}\tau_d = 50$. This leaves us with a safety margin of $1/(M\dot{\gamma}\tau) = 17$, so we do not expect too many chain crossings. Either way, this shear rate is already an order of magnitude faster than the experimental one, leaving us plenty of room for comparison with the experimental data.

In the simulation we did not reconstruct a one-to-one correspondence with either of the experimental brushes. Instead, the simulated brush density was deliberately chosen to be smaller than the experimental one, because of two reasons. First, the experimental samples, especially the Brush-short-dense, are mostly composed of the ``dry'' interior region, which would consume a lot of computing time to simulate, without resulting in any interesting effects under shear. Second, a dry and strongly stretched brush cannot be described using the same blob potential as the bulk chains. Instead, smaller blobs must be used~\cite{gay1997wetting} to ensure incompressibility which requires the total polymer density to be constant across the whole box (see Fig.~\ref{shearprofile}). Also, the brush blob size would have to shrink further as the brush collapses under shear. This introduces another complication into an already difficult system, whereas we prefer to present the absolutely simplest possible model.

\subsection{Confinement}
To confine the system between two walls, we have used the recently developed mirror-and-shift boundary conditions~\cite{korolkovas2016simulating}. Briefly, the entire system is mirrored around the $z=0$ plane and shifted by half the box length along the other two dimensions. The original system together with its mirror-shifted image is then periodically replicated in all three directions as usual, and all particles interact with their neighbours in the standard way. In other words, every particle interacts with every other particle, as well as its mirror-and-shifted images.

At this point we have a perfectly homogeneous system, and the only force driving the particles across the boundaries is the thermal noise of strength $1/\sqrt{M} \ll 1$. To block this and create the actual walls, a soft repulsive potential
\begin{equation}\label{confpotential}
U(z) = 0.05k_B T e^{-z/(2\lambda)}
\end{equation}
is applied on both sides. The range corresponds to the diameter of one blob, while the amplitude is adjusted so that the particle density in the middle of the box is equal to one. The resulting confinement force is comparatively weak, and therefore is perceived as a small perturbation to an otherwise homogeneous system. The coveted result is that the particle density (Fig.~\ref{shearprofile}) goes monotonically from zero outside the box, to one inside the box, without any overshoot or density oscillations. The wall roughness barely exceeds one blob diameter, and is about as sharp as possible. The monotonic density climb is in agreement with all of our NR measurements which strongly rule out the possibility of pronounced density oscillations near the surface.

\subsection{Grafting and shear}
To create a brush, we first generate the locations of the grafting points. For simplicity, they are arranged on a square lattice on the $z=0$ plane, plus one random number of variance $\lambda$ in all directions to make it more realistic. To ``graft'' a chain, we simply add an attractive potential between the grafting point and the central $j=J/2$ monomer:
\begin{equation}
U_{\text{graft}}(\mathbf{r}) = k_B T \cosh (r/\lambda)
\end{equation}
Half of the grafted chain $j>J/2$ is assigned to the main box and feels the same confinement potential, Eq.~\eqref{confpotential}, as all the free chains. The other half $j<J/2$ is assigned to the mirrored box, and feels the mirrored confinement $U(-z)$. This ``grafting'' technique is further explained in Ref.~\cite{korolkovas2016simulating}. In essence, at our coarse scale it is rather important to attach the central monomer and thread the chain halfway through the wall, instead of the more obvious attachment of a chain end, since this would leave a gap between the confining wall and the grafting point, and then the free chains would have a chance to unphysically cross through that gap. 

In terms of traditional end-grafted chains, our bristles have an effective length $N=256/2=128$ and there are $B=2\times8 = 16$ of them. The chain length ratio was kept to $P/N=2$ for simplicity, and is similar to the Brush-long-sparse experimental situation where the ratio is about 3. The grafting density was $0.006$ bristles per $\lambda^2 = (a\phi^{-3/4})^2$. This is about 16 times sparser than the experimental Brush-long-sparse system, but it was chosen on purpose to leave more empty space in which the brush could collapse under a broad range of shear rates, and therefore explore a wider range of conditions than possible experimentally.

The shear flow is generated by adding a Couette velocity profile:
\begin{equation}
\mathbf{v}_{\text{shear}} = \dot{\gamma} |z| \hat{\mathbf{x}}
\end{equation}
The profile is mirrored across the $z=0$ plane, so that the $j<J/2$ particles of the grafted chains also feel the shear flow in the correct direction. No slippage or shear-bands were assumed and could not easily occur in our simulation, due to the phenomenologically imposed shear flow profile. A more realistic model could better assume a constant shear stress and let the velocity profile develop instead, but we have not attempted such a simulation.

\section{Results}
\begin{figure}[ptbh!] 
\begin{subfigure}{.48\textwidth}
	\centering
		\begingroup
			\sbox0{\includegraphics{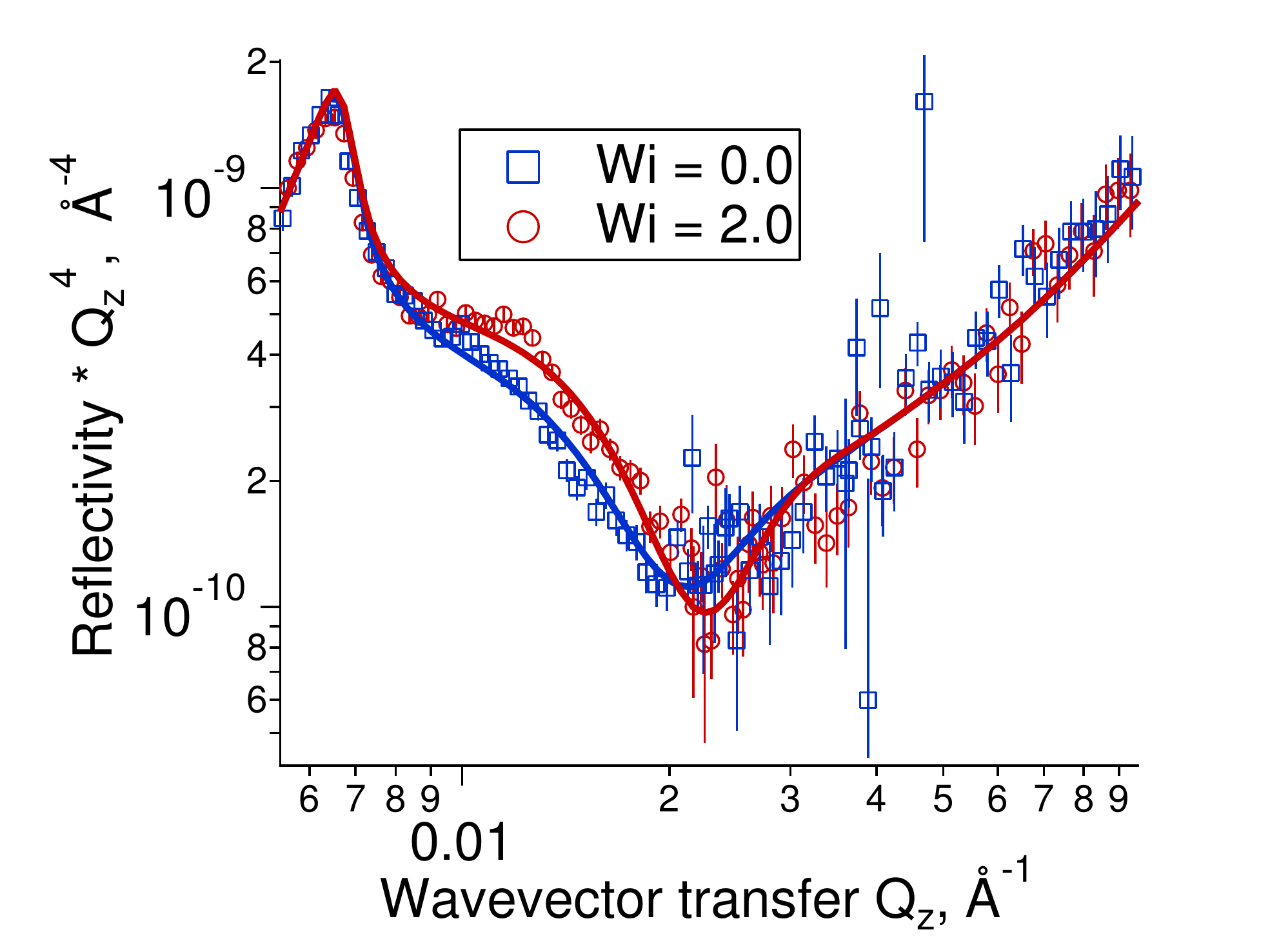}} 
			\includegraphics[clip,trim={.0\wd0} {0.0\ht0} {.08\wd0} 0,width=\linewidth]{./fignew/saclayNR.eps} 
		\endgroup
    \caption{Brush-long-sparse, NR spectrum}
    \label{saclayNR}
\end{subfigure}
\hfill
\begin{subfigure}{.48\textwidth}
	\centering
		\begingroup
			\sbox0{\includegraphics{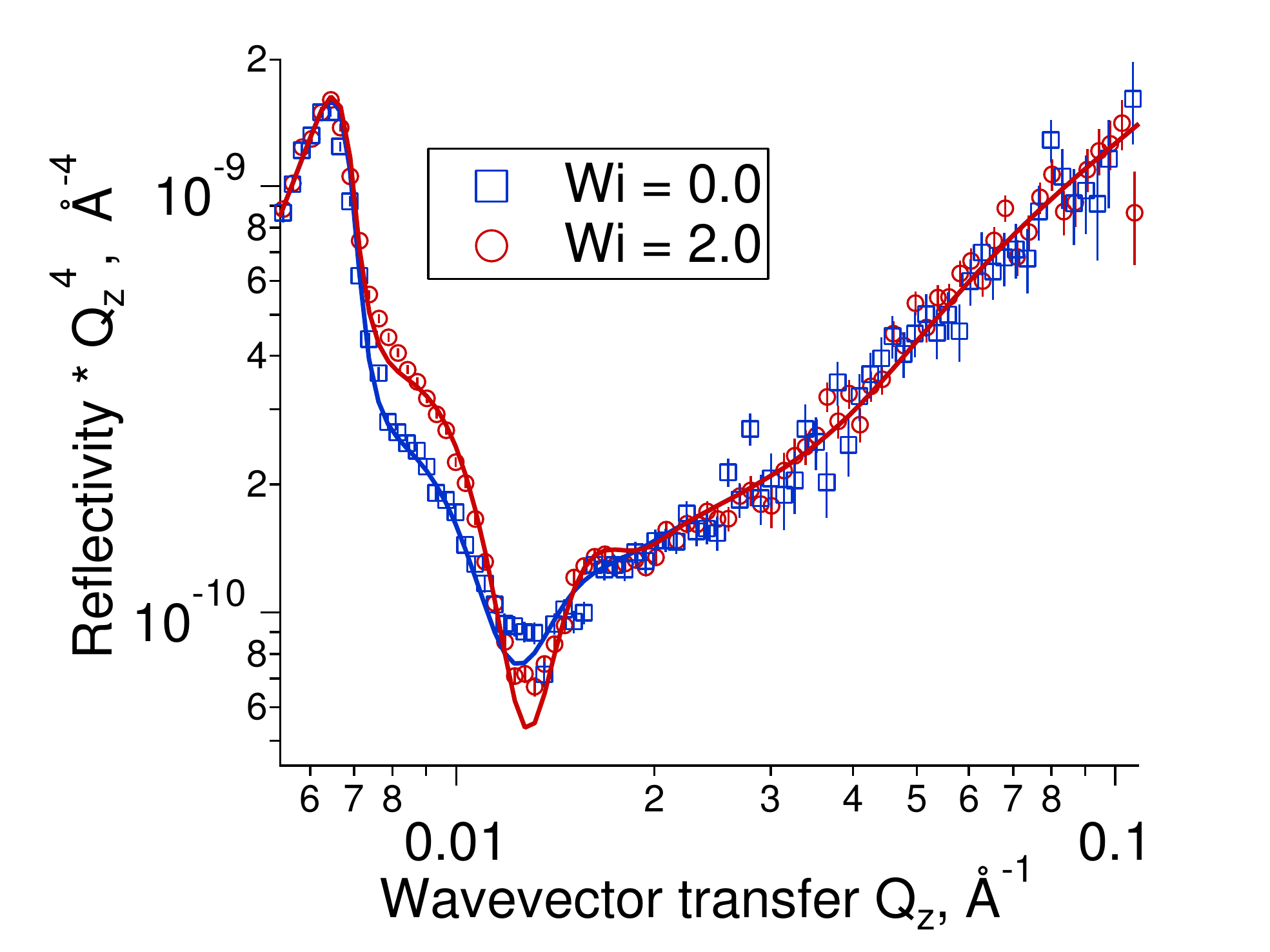}} 
			\includegraphics[clip,trim={.0\wd0} {0.0\ht0} {.08\wd0} 0,width=\linewidth]{./fignew/pragueNR.eps} 
		\endgroup
    \caption{Brush-short-dense, NR spectrum}
    \label{pragueNR}
\end{subfigure}
\vskip\baselineskip
\begin{subfigure}{0.48\textwidth} 
	\centering
		\begingroup
			\sbox0{\includegraphics{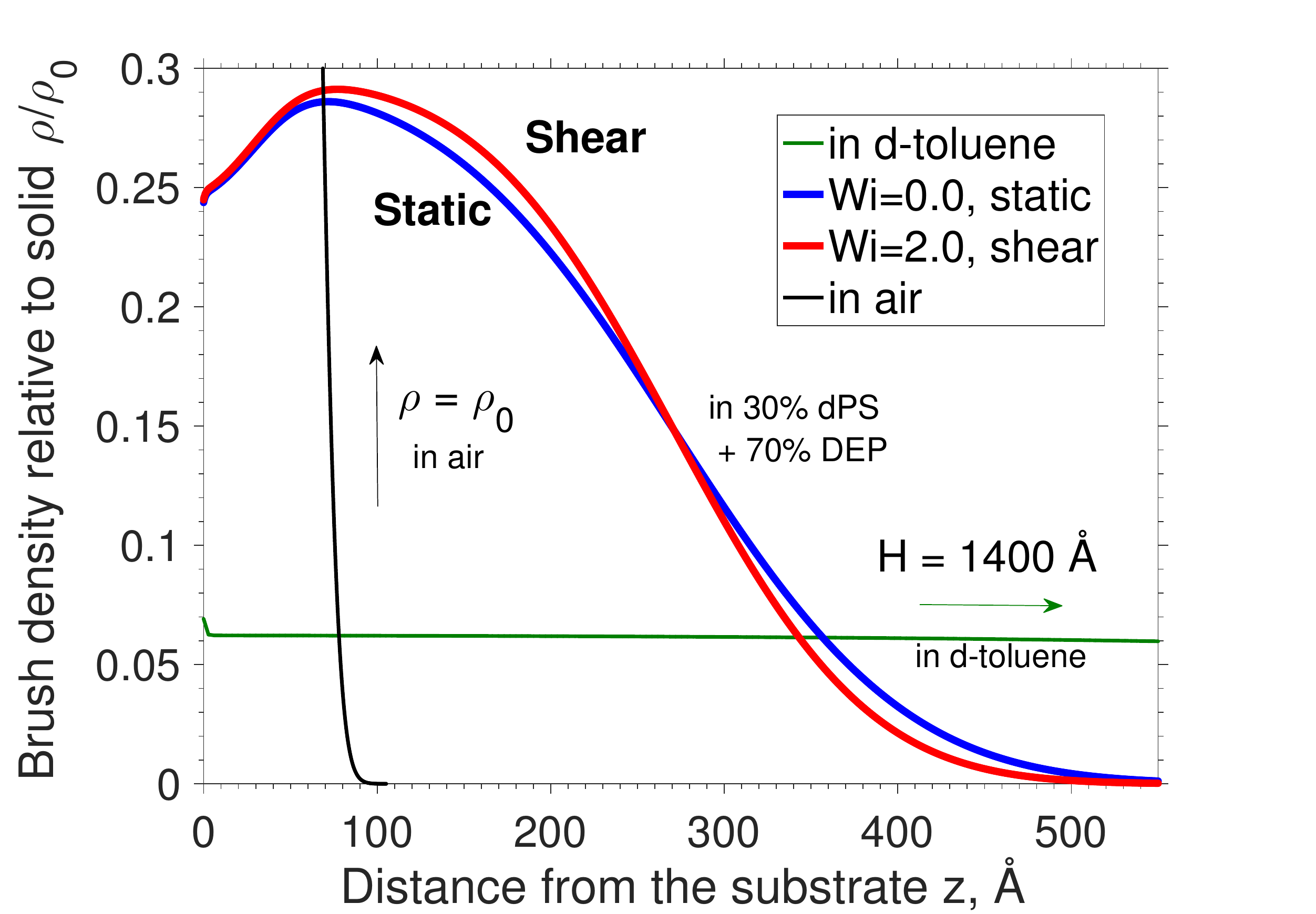}} 
			\includegraphics[clip,trim={.0\wd0} 0 {.1\wd0} 0,width=\linewidth]{./fignew/saclayprofile.eps} 
		\endgroup
    \caption{Brush-long-sparse, density profiles}
    \label{saclayprofile}
\end{subfigure}
\hfill
  \begin{subfigure}{0.48\textwidth}
	\centering
		\begingroup
			\sbox0{\includegraphics{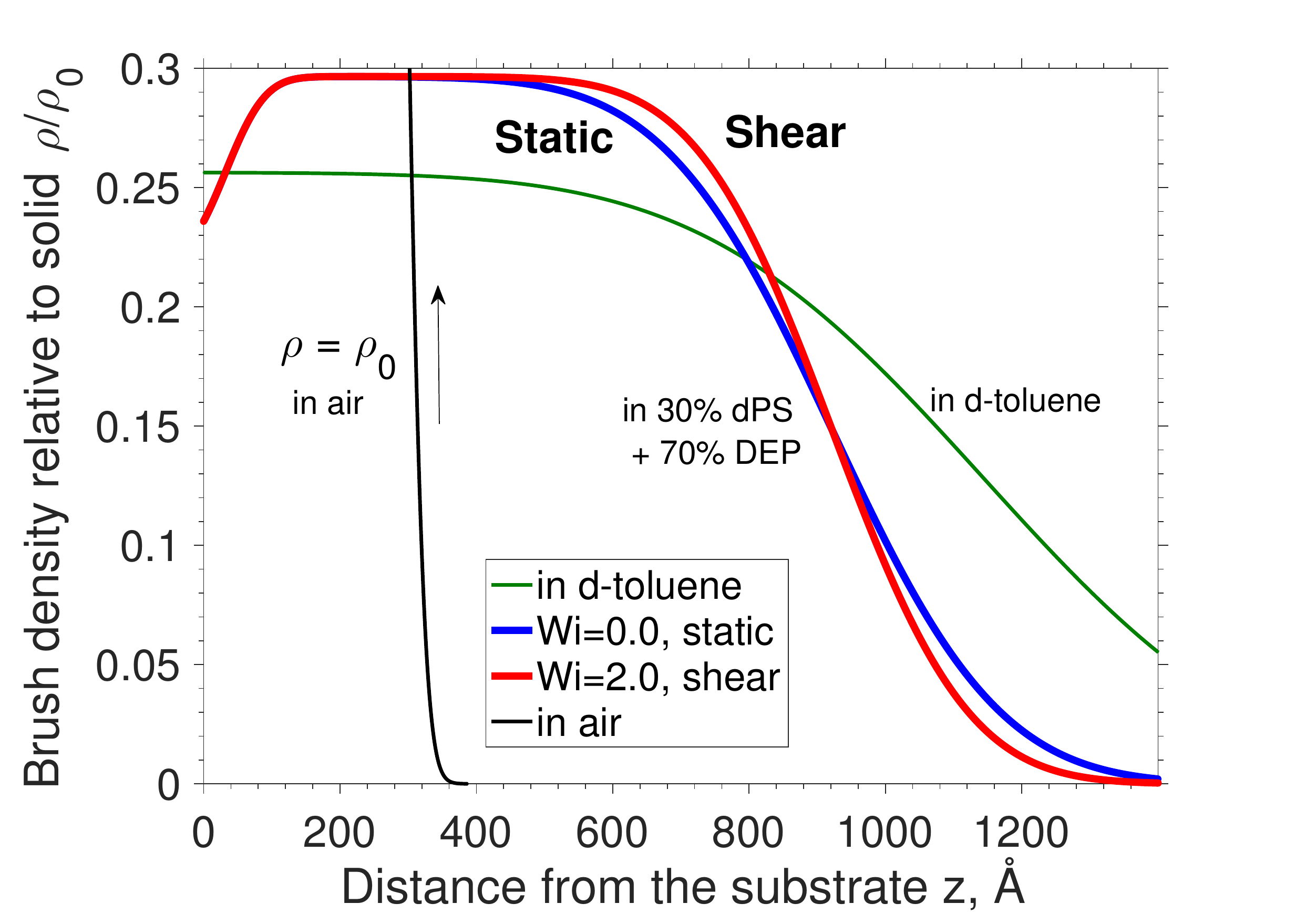}} 
			\includegraphics[clip,trim={.0\wd0} 0 {.1\wd0} 0,width=\linewidth]{./fignew/pragueprofile.eps} 
		\endgroup
    \caption{Brush-short-dense, density profiles}
    \label{pragueprofile}
\end{subfigure}
\caption{Experimentally determined brush structure. Panels a) and b) show NR data (points) and the fits (solid lines) for the two brushes in solution of 30\% dPS and 70~\% DEP. Panels c) and d) show the corresponding fitted brush density profiles (thick lines), as well as additional fits of NR measurements in air (fully collapsed), and in deuterated toluene (fully stretched). The profiles in air and d-toluene emphasize the differences between the static structure of the two brushes, whereas the relative effect of shear is about the same for both samples.}\label{expresult}
\end{figure}

Our main experimental result is shown in Fig.~\ref{expresult}. The applied shear rate $\dot{\gamma}$ is given in dimensionless Weissenberg number
\begin{equation}
\text{Wi} = \dot{\gamma} \tau_d
\end{equation}
normalized to the longest relaxation time $\tau_d$ of the bulk liquid which was measured by oscillatory rheology (see Supporting Fig.~7). The rheo-NR experiment was performed with two brushes prepared by different chemical methods which gave large differences in grafting density and molecular weight, summarized in Table~\ref{summary}: ``grafting-to'' produced a long, sparsely grafted brush (Brush-long-sparse, or Brush-LS) while ``grafting-from'' gave a shorter, denser brush (Brush-short-dense, or Brush-SD). The polymer solution was the same in both cases, $\phi = \SI{30}{\percent}$ dPS in $\SI{70}{\percent}$ DEP. The NR spectrum is displayed in panels a) and b), showing an increase of 50~\% in the reflected intensity between the static and the sheared brush. It is a strong and direct indication that the brush-bulk interface becomes sharper upon shearing. The shear was cycled on and off multiple times to demonstrate that the effect is reversible and reproducible (see Supporting Fig.~4).

To quantify the effect more precisely, we have fitted the data [solid lines in panels a) and b)] and revealed the actual brush structure in panels c) and d) respectively. The model used for the fit was verified to be consistent with information obtained by further complimentary measurements, namely the NR spectrum of the brush in air (dry, fully collapsed brush), as well as in a good solvent (maximally swollen brush) which in our case was deuterated toluene. These spectra and details about fitting are available in the SI. 

The main difference between the two brushes is their grafting density $\sigma$, defined as the number $B$ of chains per substrate area $A$, normalized by the monomer size of an effective value $a=\SI{7}{\angstrom}$ as given in Ref.~\cite{Karim1994}:
\begin{equation}
\sigma = \frac{Ba^2}{A}.
\end{equation}
Experimentally this is obtained by measuring the dry brush thickness in air
\begin{equation}
H_{\text{air}} = a \sigma N,
\end{equation}
where $N$ is the number of monomers per grafted chain. In the case of the ``grafting-from'' brush, we do not know $N$ and $\sigma$ separately. Therefore, the brush is further characterized by immersing it in a good solvent (deuterated toluene at \SI{20}{\celsius}), so the brush swells to a height~\cite{leibler1994wetting}
\begin{equation}\label{weteq}
H_{\text{good solvent}} = a N P^{-1/3} \sigma^{1/3},
\end{equation}
where $P=1$ is the length of the free chains, in this case just a single solvent molecule. The dimensionless surface coverage can then be estimated by
\begin{equation}\label{gdensity}
\sigma = \left(\frac{H_{\text{air}}}{H_{\text{good solvent}}}\right)^{3/2},
\end{equation}
comparing the dry brush thickness in air versus the thickness in a good solvent. The estimate of $\sigma$ from Eq.~\eqref{gdensity} is valid for the brushes presented here, however, it should be noted that the theoretical scaling law $3/2$ may not be exactly obeyed in general, especially for very low density brushes (mushrooms), or very short chains.

\begin{table}[htb]
\begin{tabular}{l c c}
& Brush-LS & Brush-SD\\
\hline
Chain length $N$													& 2093	&	808\\
Grafting density $\sigma$									& 0.016	& 0.16\\
\multicolumn{3}{l}{$H = $ Mean thickness (slab model), \AA} \\
\hline
In air						 				  		& 89		& 333\\ 
In d-toluene											& 1400	& 1167\\
In 30\% dPS, 70\% DEP									& 278		& 958\\
\multicolumn{3}{l}{$h = $ Brush-bulk roughness (Gaussian), \AA} \\
\hline
$\text{Wi} = 0.0$ (static)							&	105		& 194\\
$\text{Wi} = 0.5$							&	--		& 191\\
$\text{Wi} = 1.0$							&	96		& --\\
$\text{Wi} = 2.0$				&	88		& 157\\
\hline
\end{tabular}
\caption{Summary of experimental NR results.}\label{summary}
\end{table}

The summary of the brush properties determined by NR is listed in Table~\ref{summary}. There is a factor of $\sigma_{\text{SD}}/\sigma_{\text{LS}} = 10$ difference between the grafting densities of the two brushes, as well as a factor of $N_{\text{SD}}/N_{\text{LS}} = 0.4$ difference in chain length. One can better appreciate these numbers by comparing how far the Brush-LS swells in toluene (a good solvent), with respect to a more modest relative swelling of the Brush-SD, as shown in Figs.~\ref{saclayprofile} and \ref{pragueprofile}. When immersed in a \SI{30}{\percent} homopolymer solution, as opposed to a pure solvent, the excluded volume repulsion between the bristles is partially screened and the brush shrinks considerably, but is still much more swollen than the brush in air. In solution, the density profiles show two regions: 1) an interior region close to the wall where the free chains are almost completely expelled, and 2) an overlap region further out where the grafted and free chains overlap and interpenetrate.

\begin{figure}[bht]
\centering
\begin{minipage}{.48\textwidth}
	\centering
		\begingroup
			\sbox0{\includegraphics{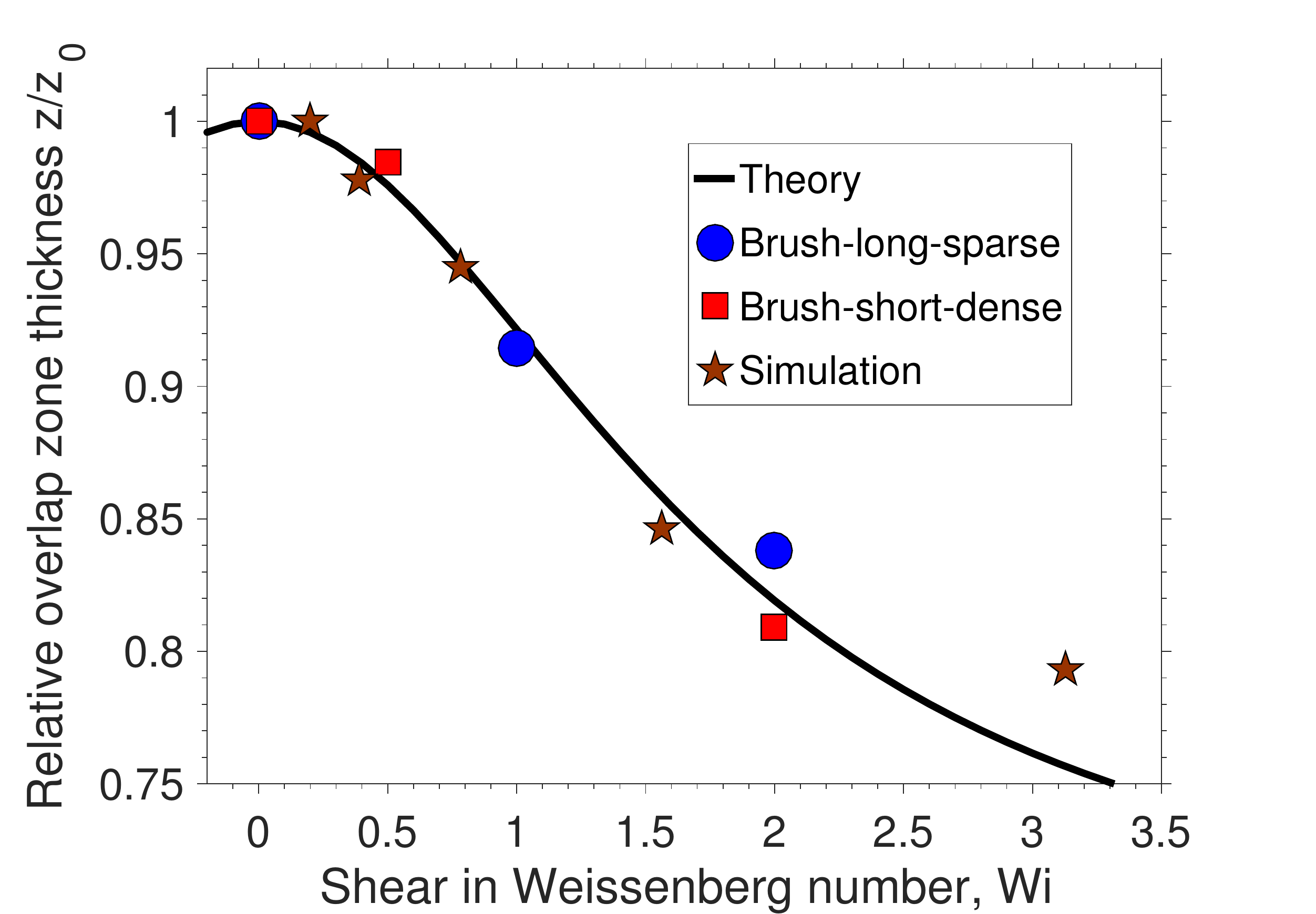}} 
			\includegraphics[clip,trim={.0\wd0} {0.0\ht0} {0.05\wd0} {0.0\ht0},width=\linewidth]{./fignew/rheight.eps} 
		\endgroup
    \caption{Brush thickness under shear, normalized to the equilibrium thickness. Only the brush-bulk overlap region is considered in this comparison. The fit (solid line) is made using Eq.~\eqref{theoryfit}.}
    \label{fig:SimNRComp}
\end{minipage}
\end{figure}

Despite the fact that the two brushes are different, the relative effect of shear on both seems to be similar, and is restricted to the overlap region. In the case of Brush-SD, its wide interior region is not affected by shear at all. Therefore, to quantify the relative change in brush structure under shear, we propose to focus on where the effect occurs and use only the mean thickness of the overlap region, which for simplicity we describe by a triangular shape
\begin{equation}
\rho(z) = \phi \left(1-\frac{z}{h}\right), \quad 0<z<h
\end{equation}
and therefore its mean thickness
\begin{equation}\label{wetthick}
\braket{z} = \frac{\int z \rho\, dz}{\int \rho\, dz} = h/3,
\end{equation}
is simply proportional to the brush-bulk roughness $h$, and does not involve the full brush thickness $H$. The relative change in the overlap thickness
\begin{equation}
\frac{\braket{z(\text{Wi})}}{\braket{z(0)}} \equiv \frac{h(\text{Wi})}{h(0)}
\end{equation}
as a function of the applied shear is plotted in Fig.~\ref{fig:SimNRComp}. Clearly, in these reduced units both brushes seem to follow a universal behaviour, within the accessible parameter range. 

To better understand the brush collapse, a series of computer simulations were performed using a previously reported algorithm for entangled polymer solutions in bulk~\cite{korolkovas2016simulation}, here extended for confined brush-bulk systems under shear flow. We have chosen one set of reasonable parameters resembling the ``grafting-to'', or Brush-LS sample, and have only varied the applied shear rate. In total, we have used $C=64$ free chains of length $P=256$ in contact with a brush containing $B=16$ grafted chains of length $N=128$. An entanglement length of $N_e = 59$ was reported in the original study~\cite{korolkovas2016simulation}, obtained using primitive path analysis~\cite{karayiannis2009combined}, leading to $Z=P/N_e = 4.3$ entanglements per chain in the bulk. The box volume is set fixed to
\begin{equation}\label{volume}
V = 2\left(\frac{4\pi}{3}\right)\lambda^3 (CP + BN)
\end{equation}
and its aspect ratio is adjusted so that the grafted chains stay far away from the opposite side of the box. To visualise the system, a smaller version was also simulated and the resulting polymer conformations were plotted in 3D, shown as insets in Fig.~\ref{static3D}.

Every simulated degree of freedom corresponds to one ``blob'', which can be mapped to the experimental system using a scaling law~\cite{deGennesScalingConcepts}
\begin{equation}\label{mapping}
N_{\text{blobs}} = \phi^{5/4} N_{\text{monomers}} .
\end{equation}
The above equation is a theoretical prediction for an ideal semi-dilute solution, up to a numerical prefactor of order one. It may require a correction if the solution is too concentrated $\phi \rightarrow 1$, which is likely for our experiment. In any case, we have made no attempt to establish an absolute one-to-one correspondence between simulation and experiment, and will content ourselves by comparing only the relative change of the brush structure as a function of the dimensionless Weissenberg number, as shown in Fig.~\ref{fig:SimNRComp}.

\begin{figure}[bth]
  \begin{subfigure}[bht]{0.55\textwidth}
	\centering
		\begingroup
			\sbox0{\includegraphics{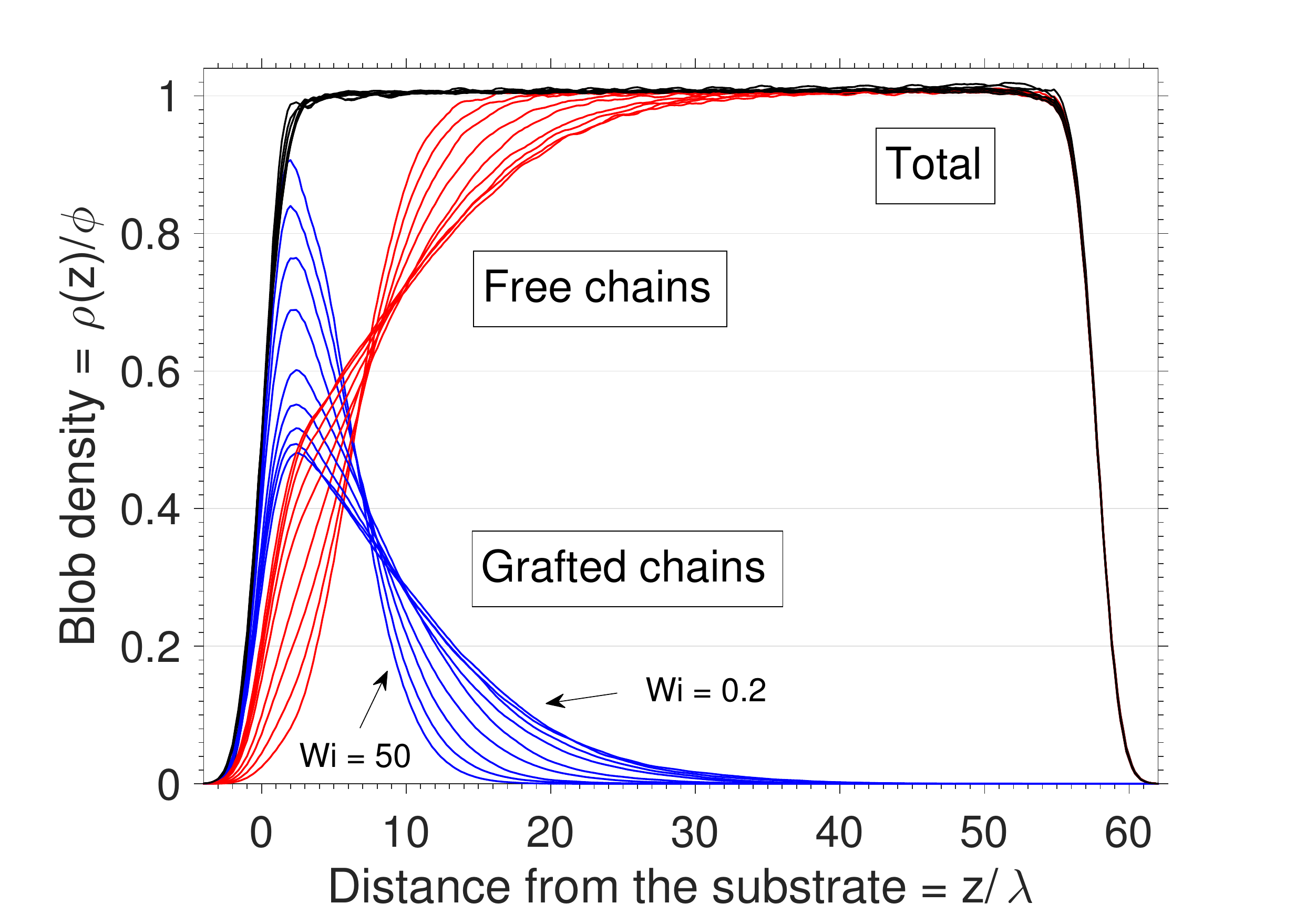}} 
			\includegraphics[clip,trim={.0\wd0} 0 {.05\wd0} 0,width=\linewidth]{./fignew/simfig/shearprofile.eps} 
		\endgroup
    \caption{Density profiles under various shear rates given in Weissenberg number $\text{Wi} = \dot{\gamma}\tau_d = 50/2^{1,\, 2,\, \ldots,\, 9}$.\newline\newline}
    \label{shearprofile}
\end{subfigure}
\begin{subfigure}{.44\textwidth}
\centering
		\begingroup
			\sbox0{\includegraphics{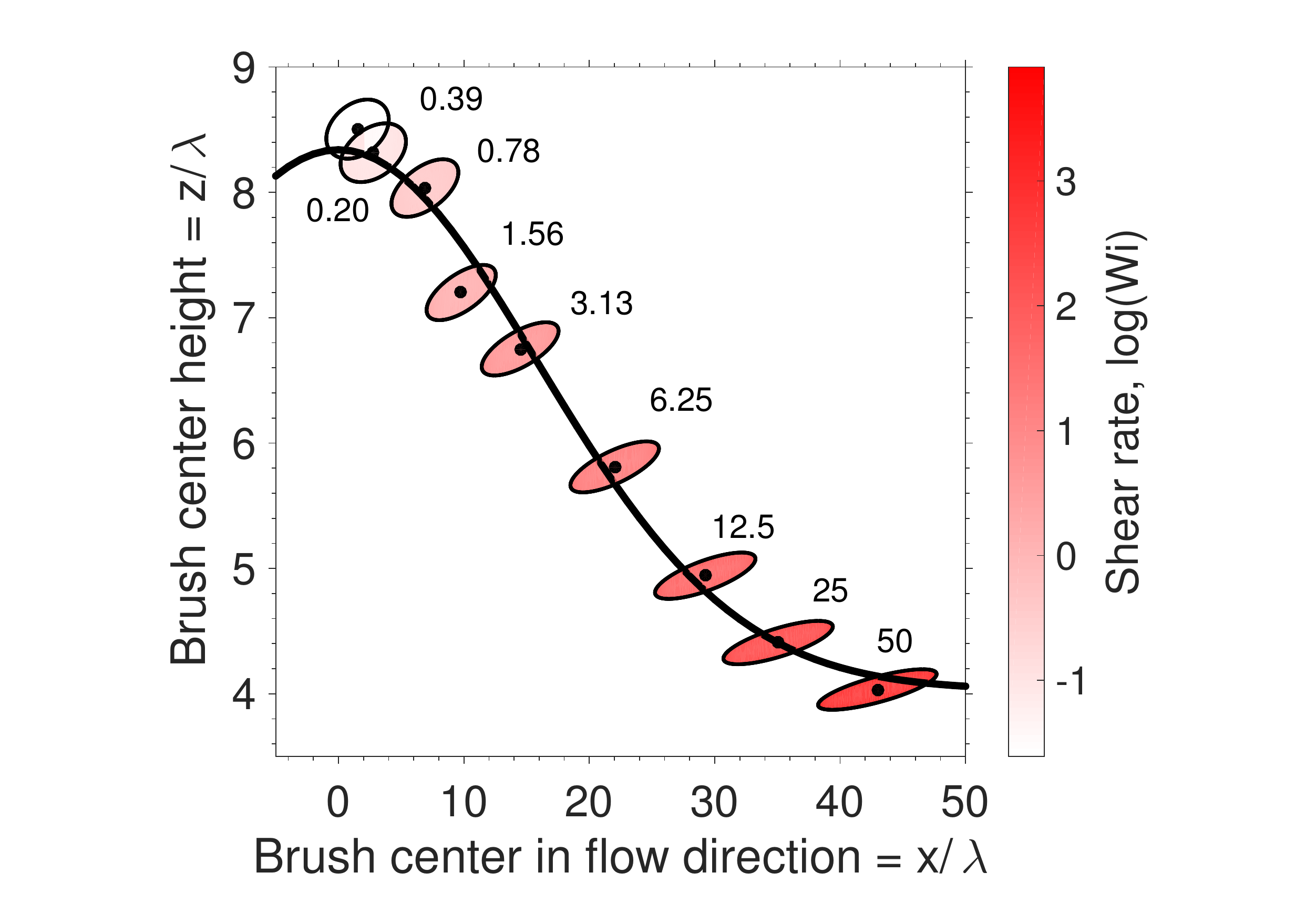}} 
			\includegraphics[clip,trim={.1\wd0} 0 {.1\wd0} 0,width=\linewidth]{./fignew/simfig/zx.eps} 
		\endgroup
    \caption{Position of the brush chain centre of mass relative to its grafting point, as defined in Eq.~\eqref{xzmean}. Ellipses show the radius of gyration scaled to the $x$-axis (see SI).}\label{zx}
		\end{subfigure}
\caption{Simulated brush structure under shear flow}\label{simresult}
\end{figure}

One advantage of simulation is that we can explore a much wider range of shear rates than possible experimentally. A shortcoming is that the computation time grows very rapidly $t\propto Z^{4.5}$ (or $t\propto Z^{3.5}+\text{overhead}$ for parallel implementations) with the number of entanglements $Z$, and systems bigger than $Z>10$ are not very practical. Keeping these considerations in mind, we simulate a lower grafting density, $\sigma_{\text{Sim}} = 0.006$ bristles per $\lambda^2$, in comparison to $\sigma_{{\text{LS}}} = 0.016$ per $a^2 = \left(\lambda \phi^{3/4}\right)^2$ for the experimental Brush-LS system. The simulated brush is thus fully overlapping with the bulk and we do not waste precious computer time to simulate any interior region which is not crucial for the brush collapse to occur. A broad range of shear rates could then be easily examined, ranging from $\text{Wi} = 0.2$ to $\text{Wi} = 50$. The resulting density profiles are shown in Fig.~\ref{shearprofile}, where the blob density is normalized to the number of blobs in the box, Eq.~\eqref{volume}. Each blob contains a $\phi$ percentage of polymer and $(1-\phi)$ percentage of solvent as mapped out by Eq.~\eqref{mapping}. The simulated density profile can be compared with the experimental one in Fig.~\ref{saclayprofile}. Even though there is a roughly $\phi^{-3/2} \sigma_{\text{LS}}/\sigma_{\text{Sim}}  = 16$-fold difference in the grafting density and about $\phi^{5/4} N_{\text{LS}}/N_{\text{Sim}}=3.6$ times difference in the chain length, the overall shape of the brush density profile and its change upon shear seem to be qualitatively similar. 

For a more quantitative comparison, we have used the definition in Eq.~\eqref{wetthick} to calculate the mean thickness of the simulated brush, and plotted the value normalized to equilibrium in Fig.~\ref{fig:SimNRComp}. When compared in terms of reduced units, there emerges a single unified trendline between the simulation and the two experiments, suggesting a common mechanism for shear-induced brush collapse in conditions where the bulk solute is entangled with the brush. Currently, we are not aware of any theoretical description which could calculate the observed brush density profiles (experimental Figs.~\ref{saclayprofile}, \ref{pragueprofile} and simulation Fig.~\ref{shearprofile}). A scaling law analysis has earlier been reported~\cite{kreer2016polymer} which roughly quantifies the brush deformation along the shear flow, but it was only intended for short, unentangled chains in which case there is no normal stress difference and hence no change in brush thickness.

Here we continue in the same scaling law spirit and propose a phenomenological explanation of our entangled brush system. At equilibrium, each bristle has a density profile $\rho(x,z)$ around its grafting point. For simplicity we will restrict ourselves to two dimensions with the $z$-direction perpendicular to the interface and the flow direction $x$. Our data indicates (Fig.~\ref{pragueprofile}) that the interior region of the brush (if present) is not affected by shear flow and therefore we will only focus on the overlap region, where the effect takes place. Its center of mass at zero shear is located at
\begin{subequations}\label{xzmean}
\begin{align}
\braket{x}_0 &= \dfrac{\int x\rho\, dx\, dz}{\int \rho\, dx\, dz} = 0\\
\braket{z}_0 &= \dfrac{\int z\rho\, dx\, dz}{\int \rho\, dx\, dz} \approx h
\end{align}
\end{subequations}
where $h$ denotes the overlap region thickness. Under a steady shear flow, the center of mass moves to some different location $\braket{x,z}$. If the shear rate is very small, one can assume phenomenologically that the displacement along the flow $\braket{x}$ is linearly proportional to the shear rate (see Supporting Fig.~9) and to the overlap thickness:
\begin{equation}
\braket{x} = (\dot{\gamma} \tau_d) \braket{z},
\end{equation}
where $\tau_d$ is the brush-bulk relaxation time, presumably governed by reptation: $\tau_d \approx \tau (P/N_e)^3 \approx 10^5 \tau$, for the simulated case. The energy penalty of the deformed brush can be estimated by
\begin{equation}
E/k_B T = \braket{x-x_0}^2 + \braket{z-z_0}^2 = (\dot{\gamma} \tau_d \braket{z})^2 + \braket{z-z_0}^2.
\end{equation}
A non-linear fluid such as ours exhibits normal stress differences and hence has a mechanism to couple the stress along various axes. The brush will therefore seek an energy minimum which can be found by solving $dE/d\braket{z} = 0$, resulting in
\begin{subequations}
\begin{align}
\braket{x} &= \left(\frac{\dot{\gamma}\tau_d}{1+(\dot{\gamma}\tau_d)^2} \right)\braket{z}_0 \label{xfit}\\
\braket{z} &= \frac{\braket{z}_0}{1+(\dot{\gamma}\tau_d)^2} \label{zfit}
\end{align}
\end{subequations}

This reasoning shows that the overall chain deformation will be smallest if the overlap thickness $\braket{z}$ shrinks below its equilibrium value, thereby avoiding some of the friction from the free chains flowing by. Of course, the brush cannot shrink to zero height, and will have to saturate to no thinner than its dry state. The simplest modification could be
\begin{equation}\label{theoryfit}
\frac{\braket{z}}{\braket{z}_0} = \frac{1-\alpha}{1+(\beta\dot{\gamma}\tau_d)^2} + \alpha
\end{equation}
with fitting parameters $\alpha=0.68$ and $\beta=0.57$, used to fit the trend in Fig.~\ref{fig:SimNRComp}.

Another great advantage of simulations is that we gain access to practically any quantity or correlation of interest, including for instance the brush center of mass displacement along the flow, $\braket{x}$, which is unavailable experimentally. We have plotted the simulated height $\braket{z}$ as a function of $\braket{x}$ for various shear rates in Fig.~\ref{zx}. In this plot both axes refer to distances, and therefore we could additionally superimpose the ellipses of inertia showing the radius of gyration of the grafted chains around their respective center of mass (more details can be found in the SI). The ellipses show that not only is the brush displaced, but it is also deformed by the shear flow, stretching in the $x$-direction, shrinking in the $z$-direction (and to a lesser extent also shrinking in the $y$-direction, see SI), and developing an anisotropic tilt, which signals the presence of shear stress~\cite{janeschitz2012polymer}. Another possible extension to Eq.~\eqref{zfit} could be a Gaussian shape:
\begin{equation}\label{theoryfit2}
\frac{\braket{z}}{\lambda} = 4.3\exp\left[-\left(\frac{\braket{x}}{22.4\lambda}\right)^2\right] + 4.0
\end{equation}
which was used to fit the simulation data in Fig.~\ref{zx}. This function also shrinks quadratically at small shear rates, $ \braket{\Delta z} \propto -\braket{x}^2$, and saturates to $\braket{z} \rightarrow \text{const.}$ at very large shear rates, but without a proper theory both Eqs.~\eqref{theoryfit} or \eqref{theoryfit2} are just guesses. Actually, the simple theoretical Eq.~\eqref{xfit} predicts that the $\braket{x}$-displacement will reach a maximum at $\dot{\gamma}\tau_d = 1$, and then slowly retract to zero. The simulation data in Fig.~\eqref{zx} clearly rules out this possibility, instead showing that the $\braket{x}$-displacement always grows monotonically and eventually saturates to some fixed value.

\section{Discussion}

At short time scales the brush behaves like a liquid, while at very long time scales like an elastic solid. The grafted chains of length $N$ relax primarily by the arm retraction mechanism~\cite{lang2016arm} $\tau_a = O(N^3 e^{N/N_e})$. This characteristic time may be further slowed ~\cite{Chenneviere2013} to $\tau_a = O(P^3 N^2 e^{N/N_e})$ during interdigitation with an entangled bulk polymer of length $P$. These very slow brush-brush relaxation processes do not couple easily to a transverse shear flow: the bristles are immobilised and cannot flow past each other. An applied shear flow only tilts the entire brush structure including its internal topological arrangements, but does not interfere with the inner brush-brush dynamics. The truly interesting coupling is between the brush and the bulk chains. These flow past each other and therefore the brush-bulk overlap region should show similar behaviours to those of the pure bulk fluid, including shear thinning and normal stress differences, expected to occur at a time scale $\tau_d = O(P^3)$ dictated by the reptation of the free chains, which should overwhelm the slower arm retraction of the brush.

The structural change observed by NR occurs almost instantly upon switching on the shear for both Brush-LS and Brush-SD, suggesting that the brush-bulk dynamics are governed by a relaxation process faster than the NR time resolution (about \SI{1}{\minute}), and therefore consistent with reptation dynamics $\tau_d \approx \SI{1}{\second}$. Overall, the brush-bulk relaxation is too fast to measure with our current setup, and the upper limit is about one minute. More information on the kinetics of the brush may be obtained in the future, using an oscillatory shear flow combined with stroboscopic NR~\cite{adlmann2015towards}. If arm retraction of the brush was to play a role, the relaxation time should be exponentially $e^{N/N_e}$ longer, and very much different for the two brushes: $\tau_{\text{LS}}/\tau_{\text{SD}} = (N_{\text{LS}}/N_{\text{SD}})^3 e^{(N_{\text{LS}}-N_{\text{SD}})/N_e} \approx 100$. In our experiment we could not detect any difference in the dynamics of the two brushes, and therefore conclude that the effect of coupling to shear flow is governed by the free chain reptation, not by the brush itself. This conclusion is corroborated by the fact that the relative brush collapse of both experimental systems and the simulation fall onto a master curve (see Fig.~\ref{fig:SimNRComp}) in spite of the different grafting densities and chain lengths of the three systems.

We emphasize that the universality of the brush collapse refers only to the brush-bulk overlap region, and does not take into account the interior brush region, which was shown here (Fig.~\ref{pragueprofile}) not to couple to the transverse shear flow, at least for the experimentally accessible shear rates. In fact, for very dense brushes the overlap region becomes too narrow to entangle with the bulk chains, in which case we could not observe any NR signal change upon shear (data not shown). We can say that the saturation parameter in Eq.~\eqref{theoryfit} becomes $\alpha = 1$, meaning that for these very dense brushes the overlap region is already fully collapsed even at shear rate $\dot{\gamma} = 0$.

One important parameter range that we have not explored is when the grafted chains are much longer than the free chains $N\gg P$, and the grafting density is sufficiently low so that more than one free chain can entangle with every grafted chain. In such a scenario the concentration of the brush is too faint to be detected by NR, at least with our present setup. Regardless of neutrons, it may happen for this system that the brush starts collapsing at $\text{Wi}\ll 1$, much sooner than the shear-thinning can erode the viscosity of the bulk liquid. If this is the case, then it may be possible~\cite{Brochard1992} that the liquid loses grip with the surface and displays a large shear-dependent surface slip. In all the cases that we studied, $N\lesssim P$, brush collapse happens at the same time as the shear-thinning in the bulk, which prevents a large slip from occurring. So far it has not been possible to characterize an appropriate $N\gg P$ system, and the surface slip question remains open.

In summary, we have used a combination of \emph{in situ} rheo-neutron reflectometry, coarse grained computer simulations and phenomenological theory to show that it is possible to engineer polymer brushes responding to shear stimuli exerted by an entangled polymer solution. At the same time we provide strong evidence that the time scale of this shear response is governed by the solution dynamics, which sets a clear limit on the tailoring of the shear-response of polymer brushes.

\section{Author contributions}
The experiment was designed by M.W., P.G., A.K., and F.R. The ``grafting-to'' brush was made by A.C. and F.R. The ``grafting-from'' brush was made by A.S.P. and C.R.E. with participation of A.K. Rheo-NR data was collected and analysed by P.G. and A.K. with participation of M.W., A.C., and F.A. Simulations and theory were performed by A.K. The manuscript was written by A.K. with the contribution of all authors. The authors declare no competing financial interests.

\section{Acknowledgements}
The authors thank Jean-Louis Barrat and Lilliane L\'eger for their invaluable comments and help. We also acknowledge the use of the Partnership for Soft Condensed Matter (PSCM) facilities and the ILL for according beam time.

\section{Data availability}
Neutron reflectometry data is available at doi.ill.fr/10.5291/ILL-DATA.9-11-1683,\\ doi.ill.fr/10.5291/ILL-DATA.9-11-1784, doi.ill.fr/10.5291/ILL-DATA.9-11-1723,\\ doi.ill.fr/10.5291/ILL-DATA.9-11-1745

Simulation algorithm is a custom written Matlab code, available upon request to the corresponding author A.K.
\newpage
\section{Supporting Information}

\section{Physical characterisation of the brushes}

\begin{suppfigure}[bht]
\begin{subfigure}{.48\textwidth}
	\centering
		\begingroup
			\sbox0{\includegraphics{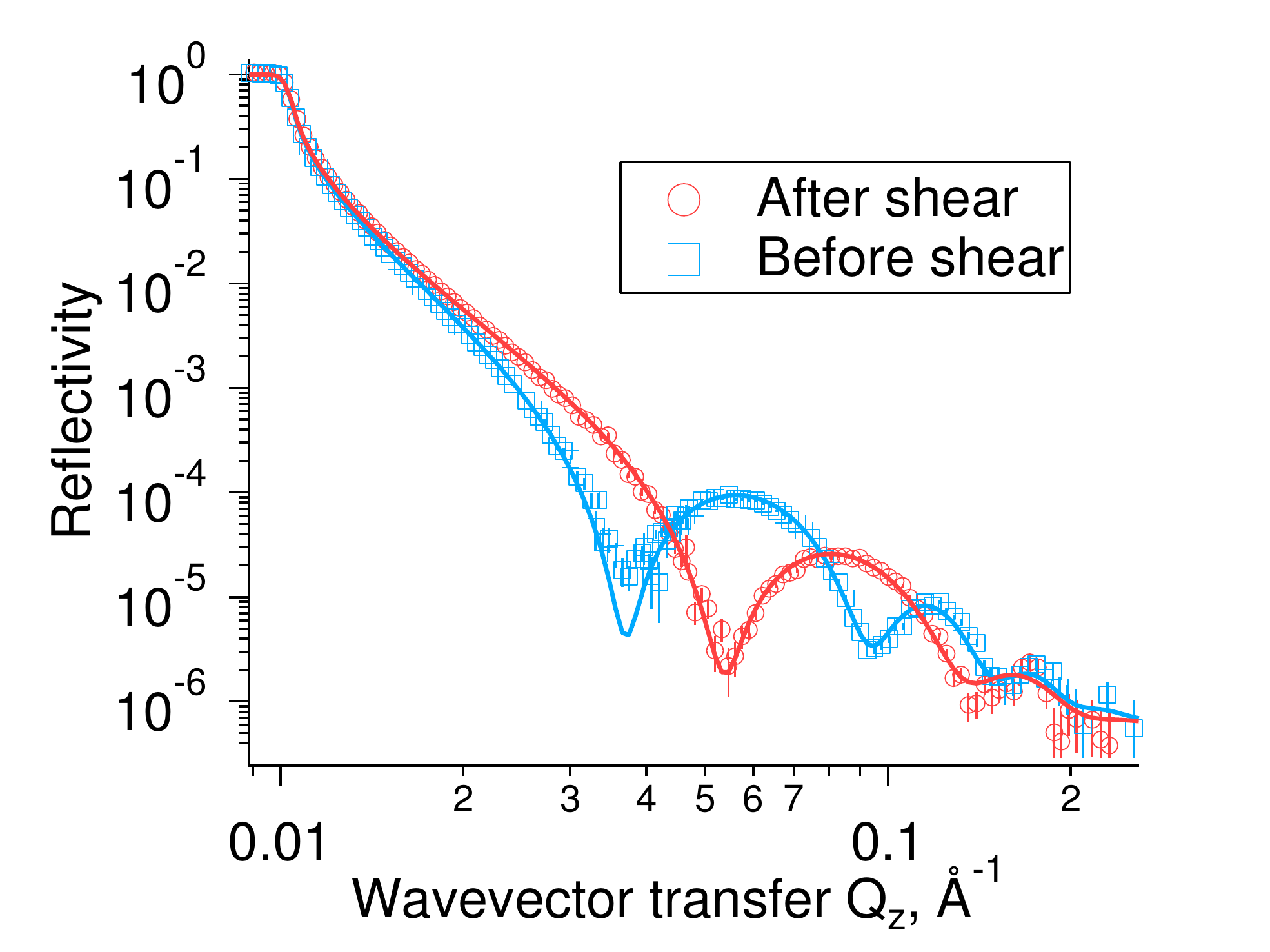}} 
			\includegraphics[clip,trim={.0\wd0} {0.0\ht0} {.1\wd0} 0,width=\linewidth]{./fignew/saclayair.eps} 
		\endgroup
    \caption{Brush-long-sparse}
    \label{saclayair}
\end{subfigure}
\begin{subfigure}{.48\textwidth}
	\centering
		\begingroup
			\sbox0{\includegraphics{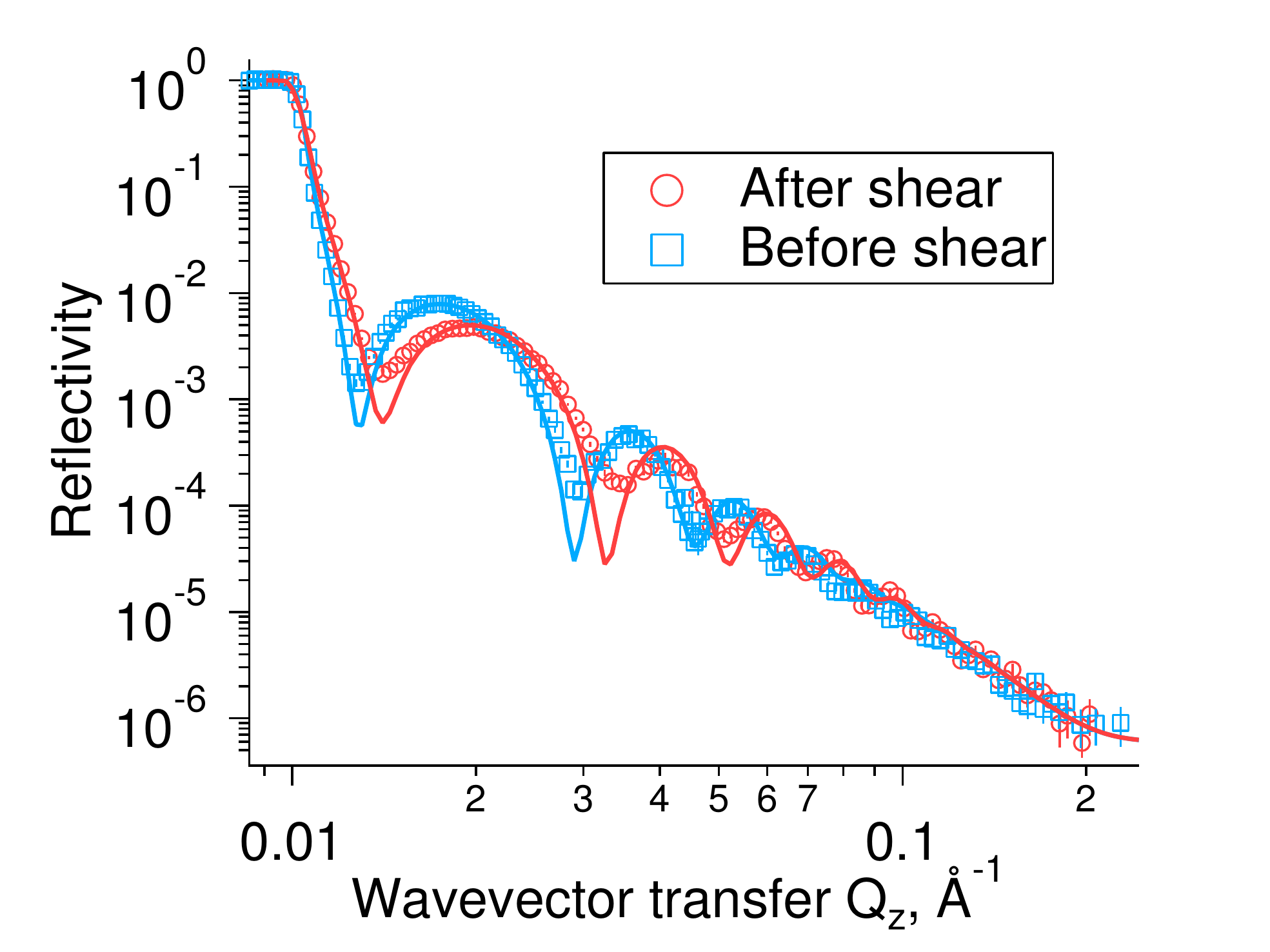}} 
			\includegraphics[clip,trim={.0\wd0} {0.0\ht0} {.1\wd0} 0,width=\linewidth]{./fignew/pragueair.eps} 
		\endgroup
    \caption{Brush-short-dense}
    \label{pragueair}
\end{subfigure}
\caption{Neutron reflectivity in air}\label{air}
\end{suppfigure}

The physical characterisation of a brush starts with a NR measurement in air shown in Supporting Fig.~\ref{air}. The dry brush thickness $H_{\text{dry}}$ can be determined by the distance between consecutive fringes $\Delta Q$:
\begin{equation}\label{fringes}
H_{\text{dry}} = \frac{2\pi}{\Delta Q}
\end{equation}
A more precise result, listed in Table~\ref{summary}, was obtained by fitting the entire spectrum using standard Motofit software, which also takes into account the native silicon dioxide layer, \SI{18}{\angstrom} and \SI{24}{\angstrom} thick, respectively.

After all the shear experiments, the brushes were thoroughly rinsed with toluene to remove any ungrafted chains. The dry air measurement was repeated again and revealed that the samples have lost 12~\% and 34~\% of their original thickness, respectively.  We presume that the brushes were gradually degraded by the strong shear stress. To simplify the remaining analysis, we will use the average thickness of before and after measurements, and assume it constant throughout the experiment.

In the dry state, the brush is fully collapsed and its height is calculated by
\begin{equation}\label{dryeq}
H_{\text{air}} = a \sigma N
\end{equation}
where $N \propto M_w$ is the number of monomers, $\sigma$ is the dimensionless grafting density, and $a$ is the size of the monomer. In the case of the ``grafting-from'' brush, we do not know $N$ and $\sigma$ separately. Therefore, the brush is further characterized by immersing it in a good solvent (deuterated toluene at \SI{20}{\celsius}), so the brush swells to a height
\begin{equation}\label{weteq2}
H_{\text{good solvent}} = a N P^{-1/3} \sigma^{1/3},
\end{equation}
where $P=1$ is the length of the free chains, in this case just a single solvent molecule. The dimensionless surface coverage can then be estimated by
\begin{equation}
\sigma = \left(\frac{H_{\text{air}}}{H_{\text{good solvent}}}\right)^{3/2}.
\end{equation}

\begin{suppfigure}[bht]
\centering
\begin{minipage}{.48\textwidth}
	\centering
		\begingroup
			\sbox0{\includegraphics{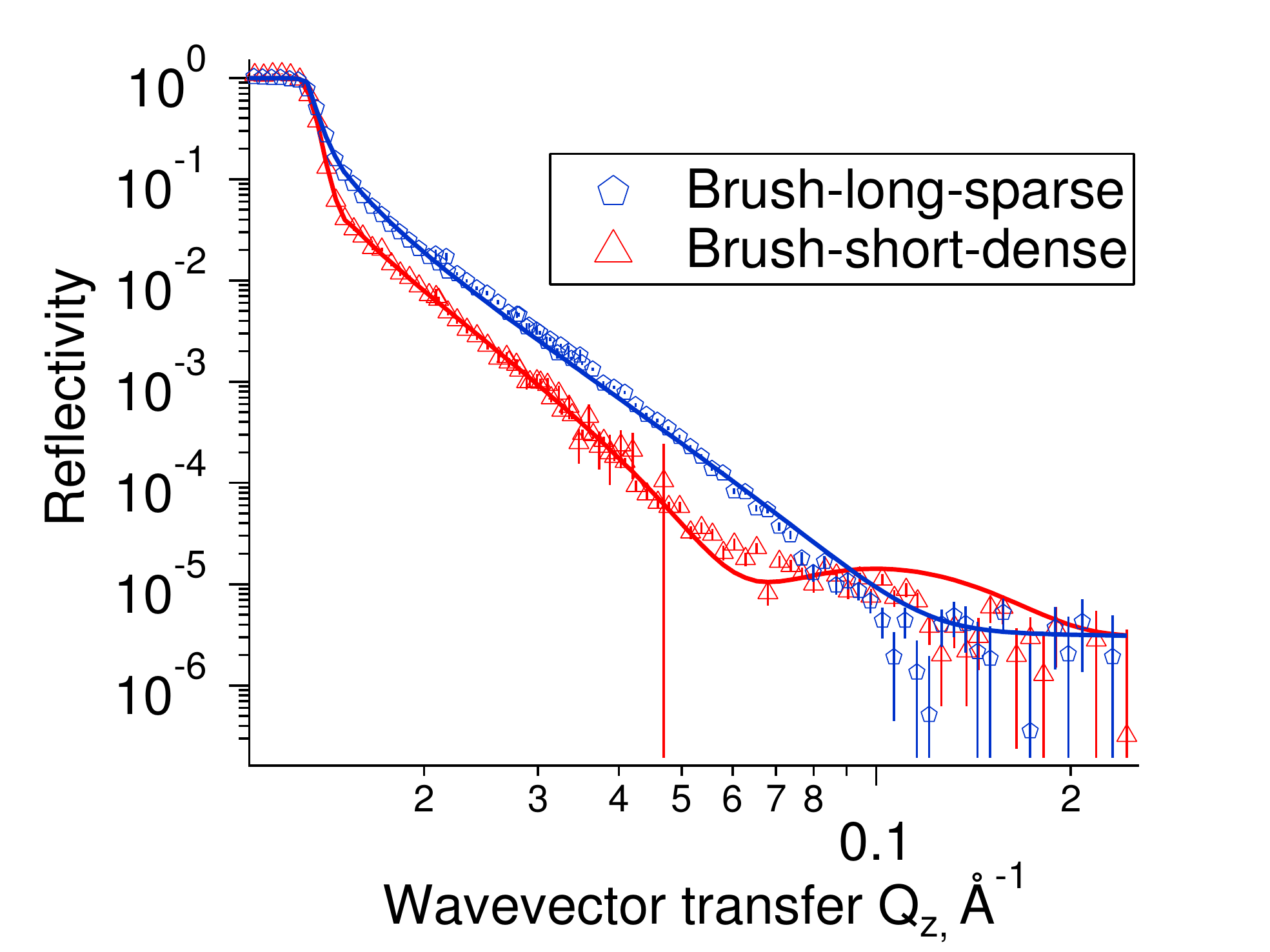}} 
			\includegraphics[clip,trim={.0\wd0} {0.0\ht0} {.1\wd0} 0,width=\linewidth]{./fignew/dtol.eps} 
		\endgroup
    \caption{Brushes swollen in deuterated toluene}
    \label{dtol}
\end{minipage}
\hfill
\centering
\begin{minipage}{.48\textwidth}
	\centering
		\begingroup
			\sbox0{\includegraphics{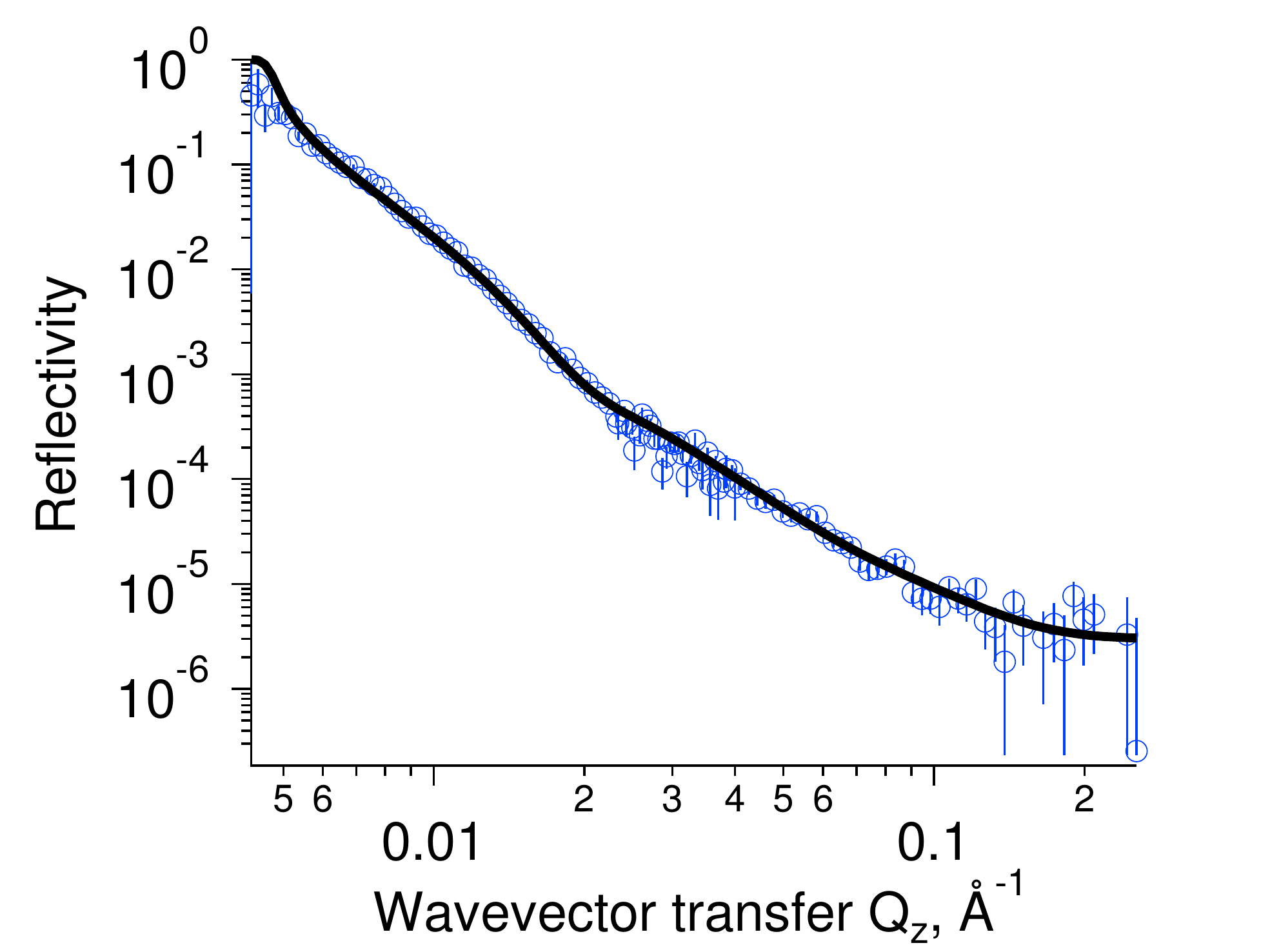}} 
			\includegraphics[clip,trim={.0\wd0} {0.0\ht0} {.1\wd0} 0,width=\linewidth]{./fignew/2ndcontrast.eps} 
		\endgroup
    \caption{Brush-long-sparse in solution of 10~\%~hPS + 20~\%~dPS + 70~\%~DEP}
    \label{2ndcontrast}
\end{minipage}
\end{suppfigure}

The corresponding NR result is shown in Supporting Fig.~\ref{dtol}. To fit the data, we have assumed that the total mass of the brush is conserved, and therefore the brush height is constrained to be
\begin{equation}
H_{\text{good solvent}} = H_{\text{air}}/\phi_g,
\end{equation}
where $0<\phi_g<\phi$ is the concentration of the grafted chains, obtained from the fitted neutron scattering length density (SLD) of the brush layer:
\begin{equation}
\underbrace{\text{SLD}[\text{hPS-dTOL}]}_{\text{from fit}} = \underbrace{\text{SLD}[\text{hPS}]}_{1.41} \phi_g + \underbrace{\text{SLD}[\text{dTOL}]}_{5.74} (1-\phi_g)
\end{equation}
In the case of the Brush-short-dense we had to include an insoluble \SI{13}{\angstrom} thick layer with $\text{SLD}=\SI{1.41e-6}{\angstrom^{-2}}$ at the base of the brush. It is attributed to the bulky ATRP initiator molecule which has a similar SLD to the h-polystyrene, but does not swell in toluene. This thickness is subtracted from the apparent dry brush thickness in air, Eq.~\eqref{fringes}, where the two species are almost indistinguishable for neutrons.

Using Eqs.~\eqref{dryeq}, \eqref{weteq2}, and the known chain length $N_{\text{LS}} = 2093$ of the Brush-long-sparse, we can estimate the unknown Brush-short-dense length:
\begin{equation}
N_{\text{SD}} = N_{\text{LS}} \left(\frac{H_{\text{SD}}}{H_{\text{LS}}}\right)_{\text{wet}}^{3/2} \left(\frac{H_{\text{LS}}}{H_{\text{SD}}}\right)_{\text{dry}}^{1/2} = 808.
\end{equation}

In the following stage, we remove the toluene by blow drying, and load the dPS-DEP solution of density $\phi = 0.3$. The repulsion between the bristles is now mostly screened by the bulk $P = 5570$ chains and the brush density profile should shrink to more of a Gaussian with height
\begin{equation}\label{gaussianeq}
H_{\text{Gauss}} = a N^{1/2}.
\end{equation}
The data in Figures~\ref{saclayNR} and \ref{pragueNR} is fitted by constraining the brush density not to exceed the bulk level of $\phi=0.3$, and maintaining the conservation of mass within reasonable bounds of 10~\%. Here we should mention that while the brush density profile changes under shear, the overall polymer concentration cannot be affected much. The relative polymer density change $\Delta \phi/\phi$ can be estimated by comparing the pressure on the cone (normal force measured at no more than $F = \SI{5}{\newton}$, spread out over a disc of radius $r=\SI{2.5}{\centi\meter}$, see also Supporting Fig.~\ref{rheology2}), against the osmotic pressure $\Pi$ of the polymer solution:
\begin{equation}
\frac{\Delta \phi}{\phi} \approx \frac{\Delta \Pi}{\Pi} \approx \frac{F \lambda^3}{\pi r^2 k_B T} \approx 0.001,
\end{equation}
where we have estimated the blob size at $\lambda \approx a/\phi^{3/4} \approx \SI{7}{\angstrom}$. Thus we can see that the shear-induced concentration change is minuscule and can be disregarded.

The NR fits are considerably improved if we allow for a slightly depleted ($\phi_d \approx 0.24$) thin layer at the base of the brush. This assumption is verified by measuring the brush in a different contrast, consisting of 10~\% hPS and 20~\% dPS solution in DEP, shown in Supporting Fig.~\ref{2ndcontrast} for the Brush-long-sparse. The fit was produced assuming the same brush structure, but different SLD weights. The only small discrepancy was that in the second contrast, the thin depletion layer had a smaller density of $\phi_g = 0.18$. The final brush height reported in Table~\ref{summary} is just the sum of the depleted and the main brush layers.

It is apparent that the simple Eq.~\eqref{gaussianeq} is not obeyed by our samples: $(H_{\text{SD}}/H_{\text{LS}} = 3.4) \neq (\sqrt{N_{\text{SD}}/N_{\text{LS}}} = 0.6)$. Therefore, the brushes cannot be considered Gaussian, especially the denser Brush-short-dense. In the case of polymer melt, one could interpolate Eqs.~\eqref{dryeq}, \eqref{weteq2} and \eqref{gaussianeq} with this function:
\begin{align}
H &= a N^{1/2} \left(1 + \frac{N^{1/2} \sigma^{1/3}}{P^{1/3}} \left( 1 + \sigma^{2/3} P^{1/3}\right) \right) \\
&= a N^{1/2} \left(1 + \left(\frac{N}{P}\right)^{1/2} \left(P\sigma^2\right)^{1/6} \left(1 + \left(P\sigma^2\right)^{1/3} \right) \right)
\end{align}
In a semi-dilute solution of density $(\phi^{*}\approx 0.03)<(\phi=0.3)<(\phi^{**}\approx 0.5)$, the above equation can be extended by a mapping from the blob theory: $N \rightarrow \phi^{5/4}N$, $P \rightarrow \phi^{5/4}P$, $a \rightarrow \lambda = a \phi^{-3/4}$ and $\sigma \rightarrow  (\lambda/a)^2 \sigma = \phi^{-3/2} \sigma$. The brush height in the general case can then be interpolated as
\begin{equation}\label{general}
H = a \left(\frac{N}{\phi^{1/4}}\right)^{1/2} \left(1 + \alpha \left(\frac{N}{P}\right)^{1/2} \beta \left(\frac{P\sigma^2}{\phi^{7/4}}\right)^{1/6} \left(1 + \beta^2 \left(\frac{P\sigma^2}{\phi^{7/4}}\right)^{1/3} \right) \right).
\end{equation}
The above equation is fitted using our six height measurements, to obtain the three fitting parameters: $a = \SI{1.19}{\angstrom}$, $\alpha = 1.95$, and $\beta=0.94$. The locations of the various brush states are indicated in the phase diagram, Supporting Fig.~\eqref{phasediagram}. It turns out that the Brush-short-dense in \SI{30}{\percent} dPS solution is best described as dry, meaning that the free dPS chains are largely expelled from the brush. The Brush-long-sparse is quite close to the triple cross-over between dry, Gaussian and stretched, but leans more to the dry side.


\begin{suppfigure}[tbh]
    \begin{minipage}{.6\textwidth}
	\centering
		\begingroup
			\sbox0{\includegraphics{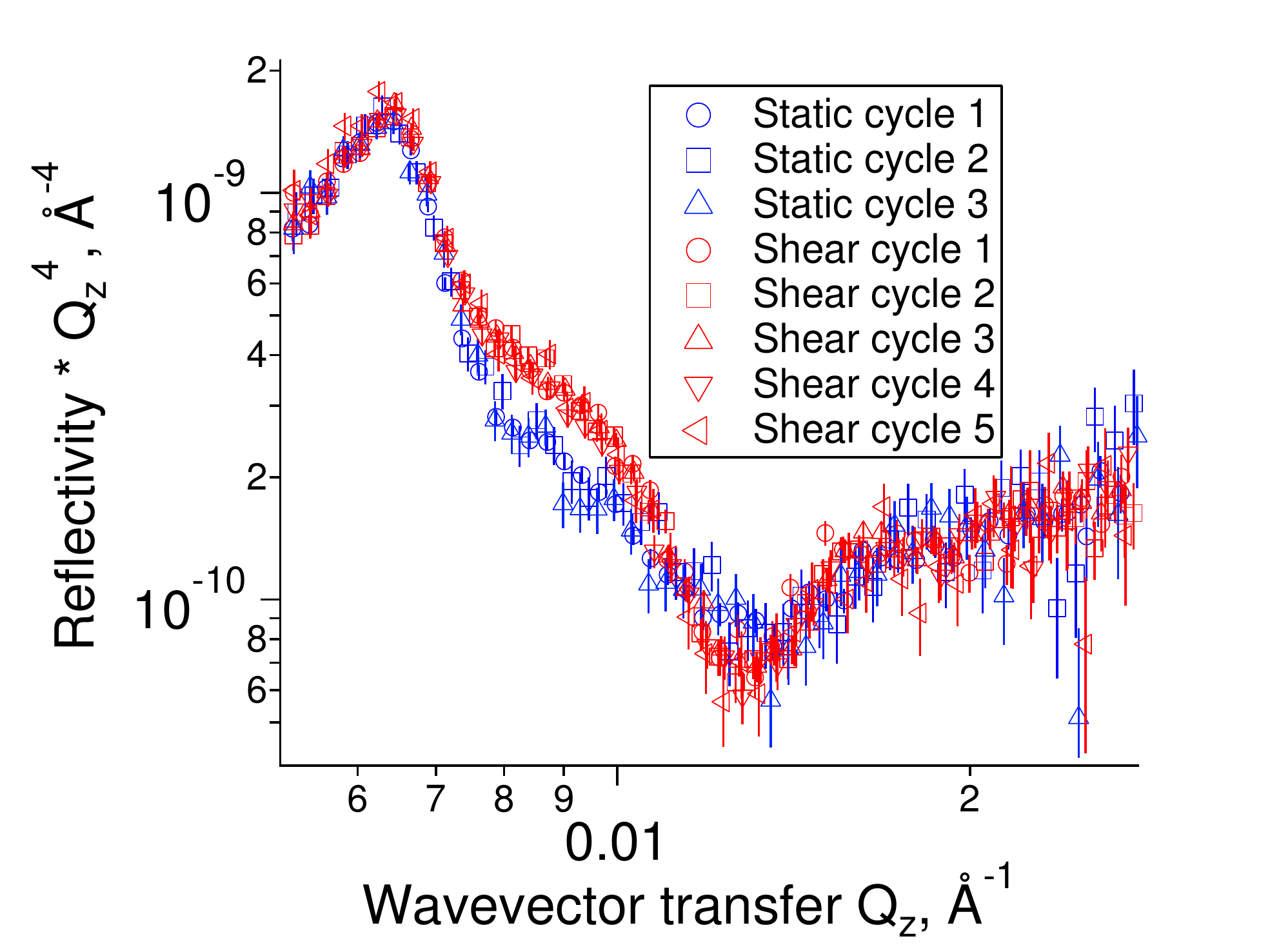}} 
    			\includegraphics[clip,trim={.0\wd0} {.0\ht0} {.0\wd0} 0,width=\linewidth]{./fignew/reproducibility.eps} 
		\endgroup
    \caption{Reproducibility of the brush collapse (Brush-short-dense sample)}
    \label{reproducibility}
\end{minipage}
\end{suppfigure}

\begin{suppfigure}[bht]
\centering
		\begin{minipage}{.6\textwidth}
	\centering
		\begingroup
			\sbox0{\includegraphics{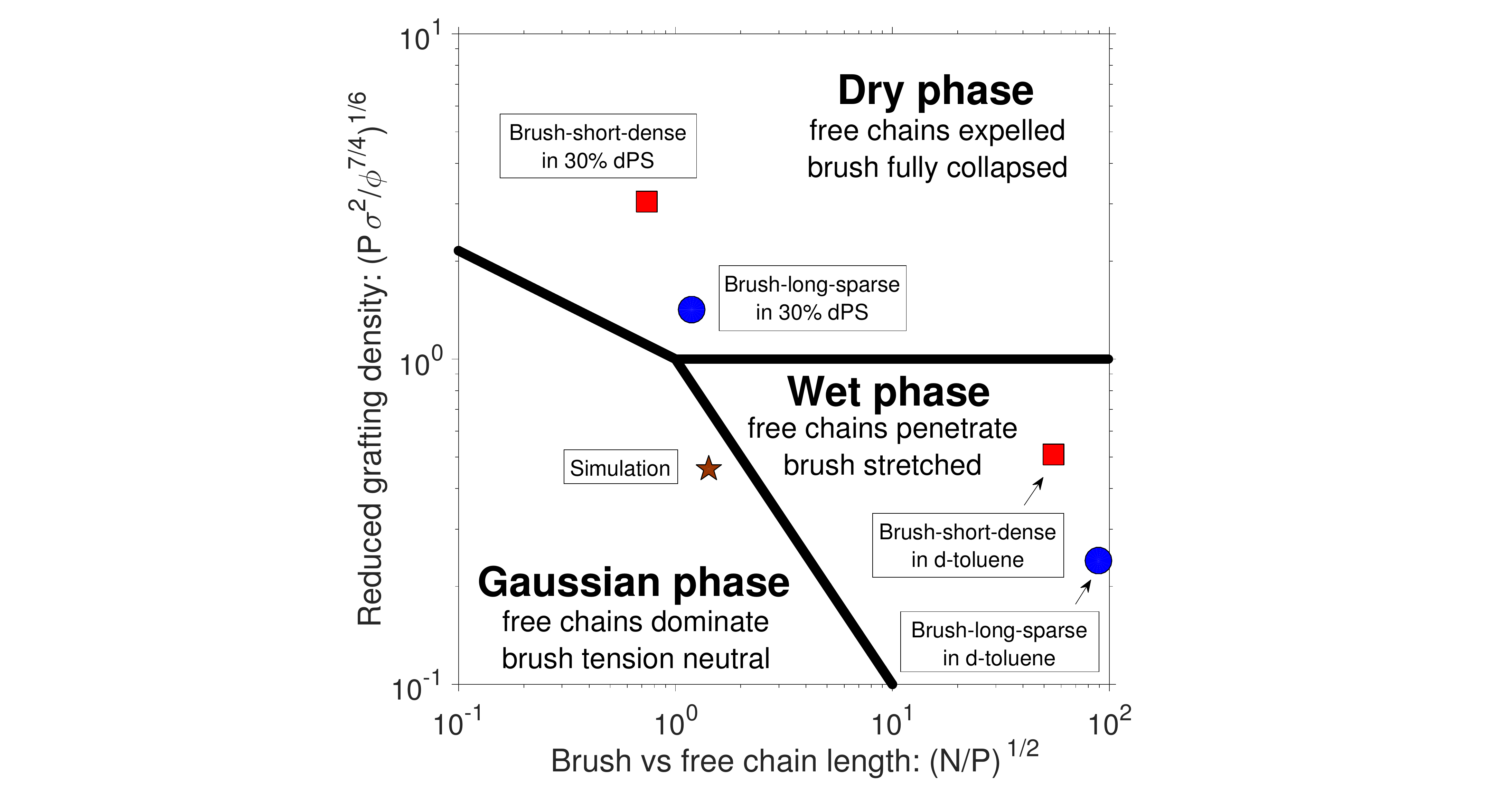}} 
			\includegraphics[clip,trim={.2\wd0} 0 {.25\wd0} 0,width=\linewidth]{./fignew/phasediagram.eps} 
		\endgroup
		\caption{Thermodynamic phase diagram indicating the various brush states.}
    \label{phasediagram}
\end{minipage}
\end{suppfigure}


\section{Confirmation of Chemical Structure of PS Brushes via FTIR spectroscopy}
\begin{suppfigure}[bht]
\centering
\begin{minipage}{.48\textwidth}
	\centering
		\begingroup
			\sbox0{\includegraphics{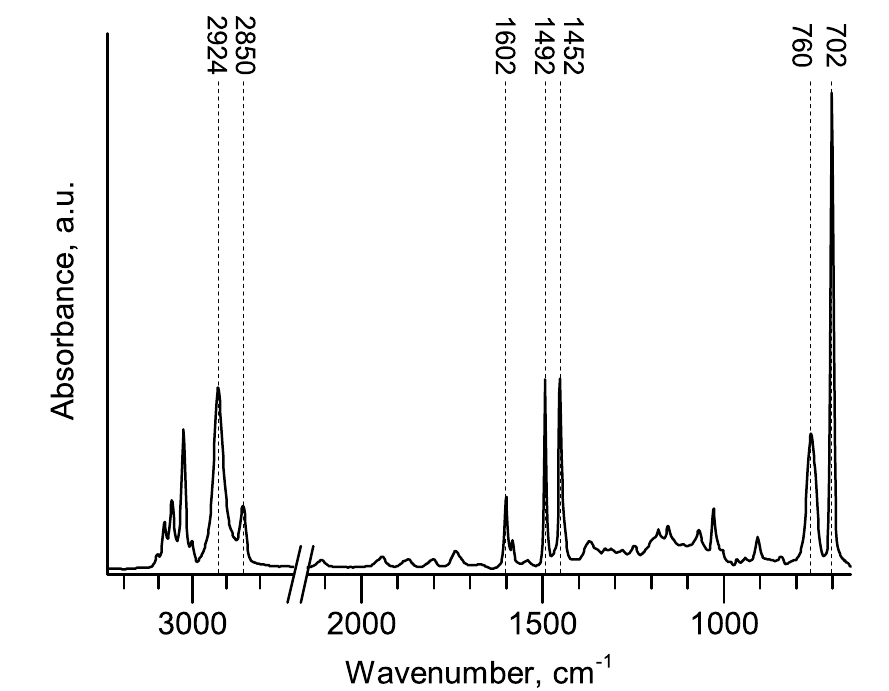}} 
			\includegraphics[clip,trim={.0\wd0} 0 {.0\wd0} 0,width=\linewidth]{./fignew/FTIR.eps} 
		\endgroup
    \caption{FTIR spectrum of ``grafted-from'' PS brush}
    \label{FTIR}
\end{minipage}
\end{suppfigure}

The chemical structure of the ``grafted-from'' PS Brush-short-dense was confirmed via Fourier-transform infrared spectroscopy in attenuated total reflectance mode (ATR-FTIR) employing a Nexus 870 spectrometer (Nicolet, Czech Republic) equipped with a VariGATR ATR accessory (Harrick Scientific Products, USA). Measurements were performed using 256 scans at \SI{4}{\centi\meter^{-1}} on a sample prepared in parallel with the sample for neutron reflectometry.
The spectrum obtained is shown in Supporting Fig.~\ref{FTIR}, displaying features characteristic for PS. The CH${}_2$ stretching modes of the polymer backbone are observed at \SI{2924}{\centi\meter^{-1}} (symmetric) and \SI{2850}{\centi\meter^{-1}} (asymmetric) while the band at \SI{1452}{\centi\meter^{-1}} arises mostly from the CH${}_2$ bending mode. A series of 5 weak bands between 1945 and \SI{1672}{\centi\meter^{-1}} are the result of combination vibrations of the aromatic ring and the bands appearing at 1602 and \SI{1492}{\centi\meter^{-1}} arise from in-plane ring vibrations. The strong bands at 702 and \SI{760}{\centi\meter^{-1}} correspond to out-of-plane ring deformations.

\section{Rheological characterisation}
\begin{suppfigure}[bht]
\centering
\begin{minipage}{.46\textwidth}
	\centering
		\begingroup
			\sbox0{\includegraphics{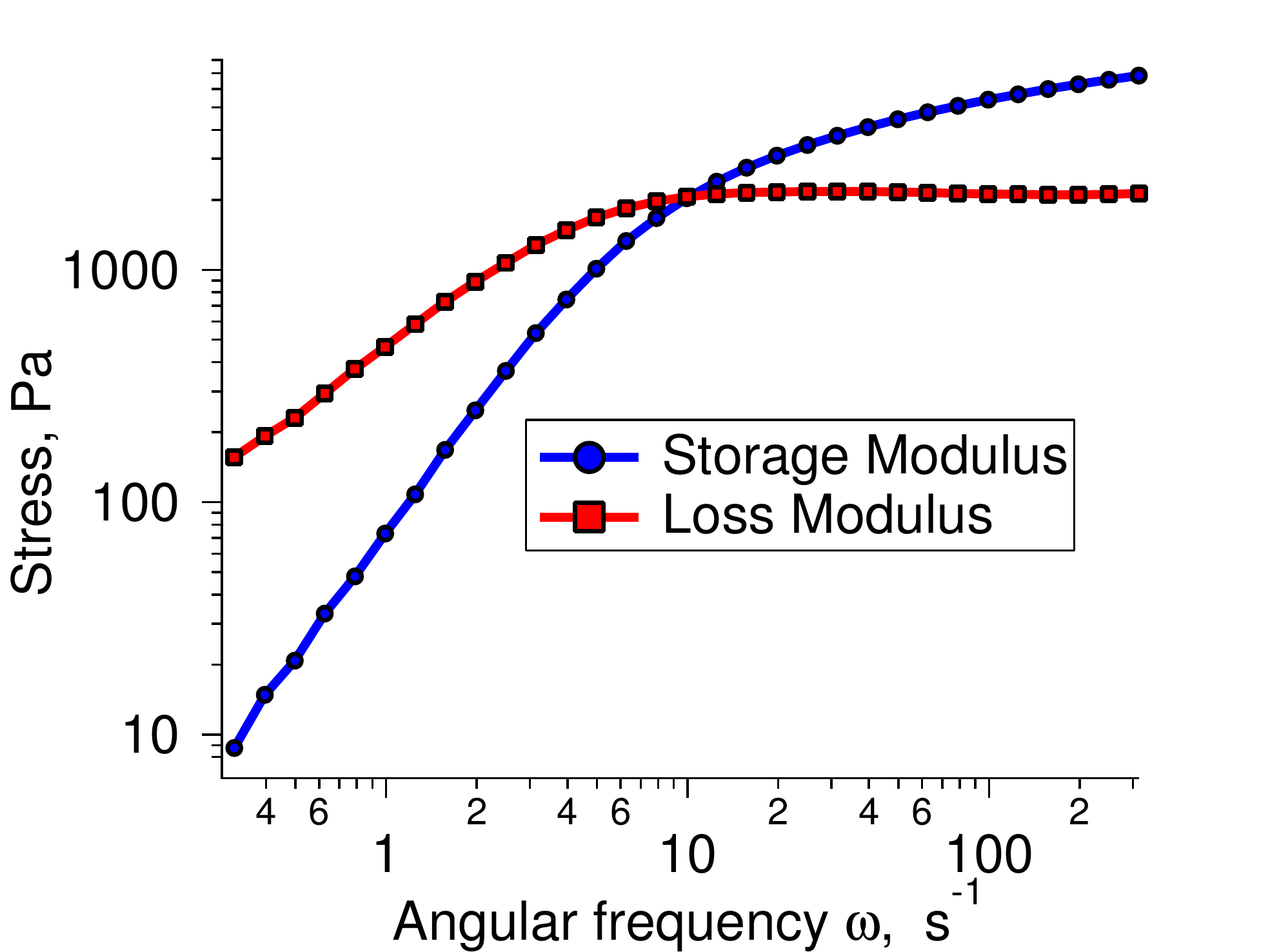}} 
			\includegraphics[clip,trim={.0\wd0} {0.0\ht0} {.0\wd0} 0,width=\linewidth]{./fignew/rheology.eps} 
		\endgroup
    \caption{Storage and loss moduli of 30~\% dPS solution in DEP at 5~\% strain amplitude.}
    \label{rheology}
\end{minipage}
\hfill
\begin{minipage}{.51\textwidth}
	\centering
		\begingroup
			\sbox0{\includegraphics{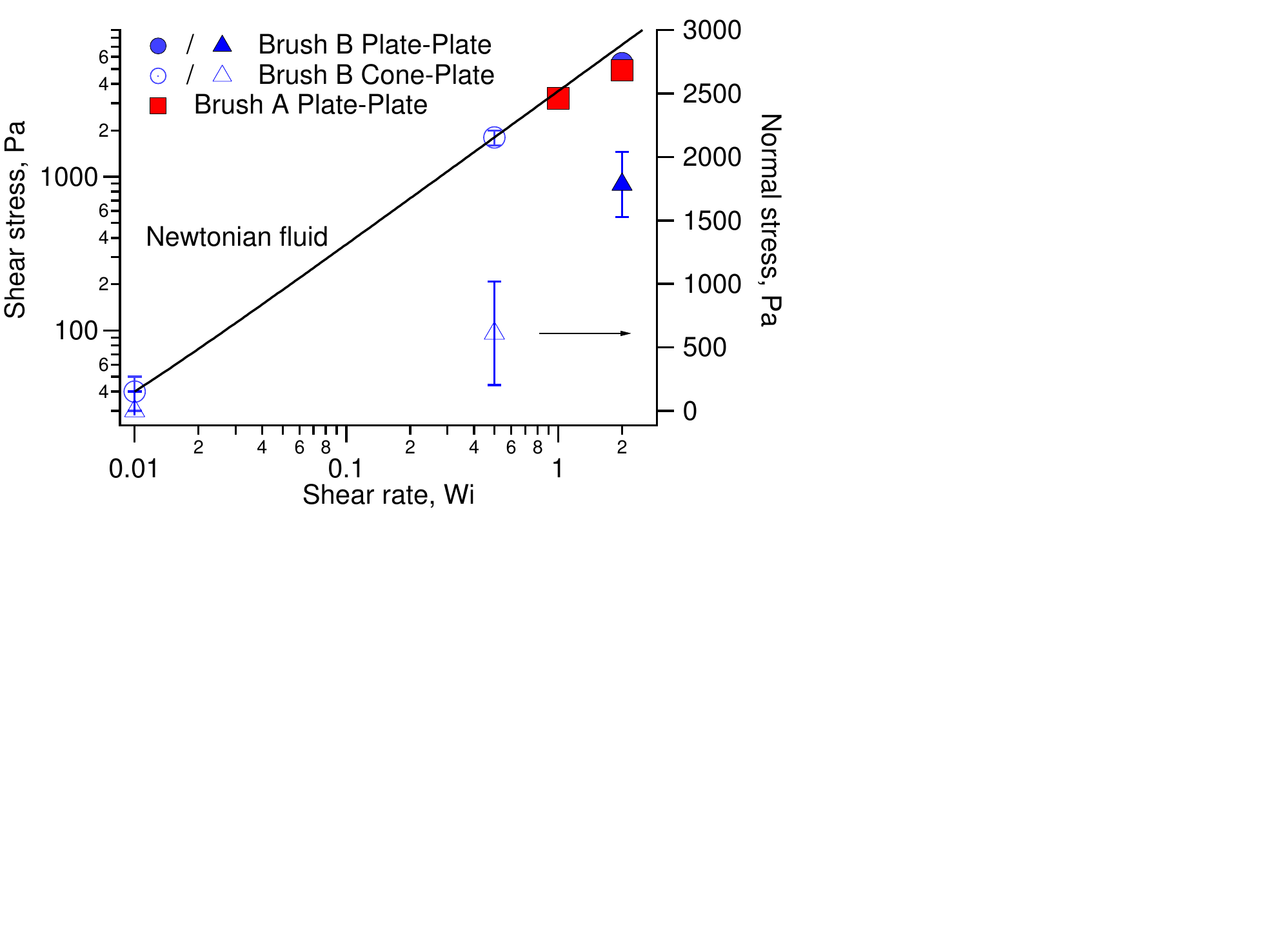}} 
			\includegraphics[clip,trim={.0\wd0} {0.45\ht0} {.35\wd0} 0,width=\linewidth]{./fignew/rheology2.eps} 
		\endgroup
    \caption{Steady-state shear and normal stresses as recorded during the \emph{in situ} rheo-NR measurements}
    \label{rheology2}
\end{minipage}
\end{suppfigure}
The viscoelastic properties of the bulk dPS solution were characterised by measuring the storage $G'(\omega)$ and loss $G''(\omega)$ moduli against the angular frequency $\omega$ of the applied shear rate, shown in Supporting Fig.~\ref{rheology}. The curves are typical for a viscoelastic fluid, and their cross-over at frequency $\omega^* = \SI{10}{\per\second}$ determines the longest relaxation time of the bulk polymer, in this case $\tau_d = 1/\omega^* = \SI{0.1}{\second}$. The shear stress as well as the normal stress measured during the NR data acquisition is also shown in Supporting Fig.~\ref{rheology2}. It is clear that that after $\text{Wi}\gtrsim 1$ we enter into a non-Newtonian shear-thinning regime. The outwards normal stress on the cone also starts rapidly increasing at this point. These rheological observations coincide closely with the onset of the brush collapse as measured by NR.

\section{Simulated brush structural analysis}
The density profiles in Fig.~\ref{shearprofile} are histograms obtained by counting the number of particles in small bins and averaging over about $10\tau_d$ time steps. Further information on the brush structure is obtained by calculating the average position of the center of mass for the average bristle, with respect to its grafting point. The mean position is shown by black points in Fig.~\ref{zx}. We have also calculated the components of the inertia tensor:
\begin{equation}
R_{\alpha \beta} = \braket{(R-R_0)_{\alpha} (R-R_0)_{\beta}},
\end{equation}
where $\mathbf{R}_0$ is the instantaneous position of the center of mass, and the average is taken over all the $j$-particles. At equilibrium, the sum of diagonals $R_{xx}+R_{yy}+R_{zz} = R_g^2$ is known as the radius of gyration. Under shear, there will also be a non-zero off-diagonal component $R_{xz}$ and the inertia tensor can be described as an ellipse. To quantify its shape, we must solve the diagonalization problem:
\begin{align}
(R_{xx}-A)\cos \alpha	+	R_{xz}\sin \alpha &= 0\\
R_{xz}\cos \alpha	+	(R_{zz}-A)\sin \alpha &= 0
\end{align}
The solution is the tilt angle:
\begin{equation}
\tan 2 \alpha = \frac{2R_{xz}}{R_{xx} - R_{zz}}
\end{equation}
and the principal axes of inertia:
\begin{equation}
R_{1,2} = \frac{1}{2}\left[(R_{xx}+R_{zz}) \pm \sqrt{(R_{xx}-R_{zz})^2 + 4R_{xz}^2}\,\right]
\end{equation}
The resulting ellipse is drawn around the position of the brush center in Fig.~\ref{zx}. The ellipse dimensions are scaled to the values on the $x$-axis, which has a ratio of 10:1 with respect to the $y$-axis. The shear rate in Weissenberg number is denoted by the color inside each ellipse and also the number next to it. The full dataset is further shown in Supporting Fig.~\ref{Rg} in a plain format.

\begin{suppfigure}
	\centering
		\begingroup
			\sbox0{\includegraphics{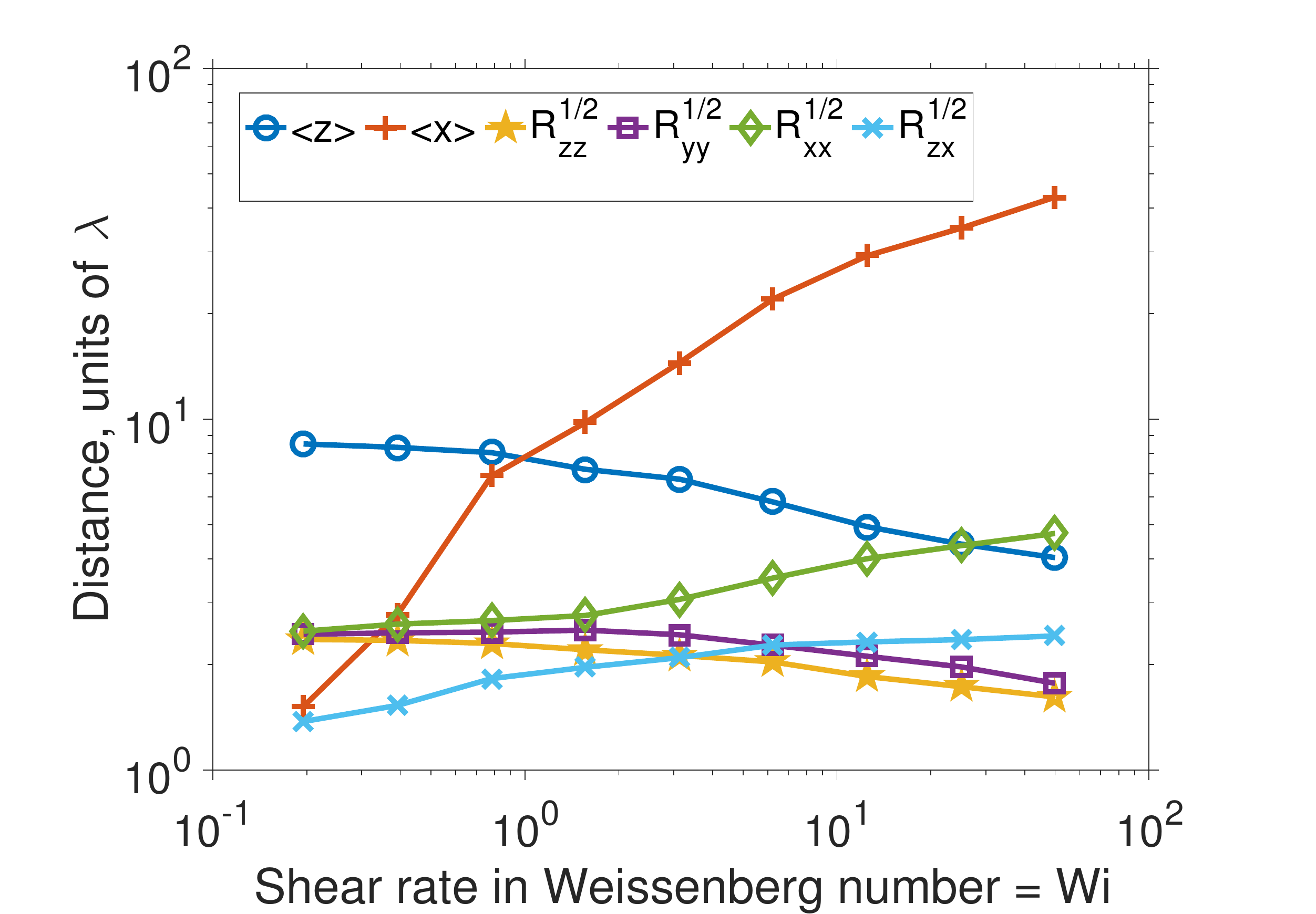}} 
			\includegraphics[clip,trim={.05\wd0} 0 {.1\wd0} 0,width=0.55\linewidth]{./fignew/simfig/Rg.eps} 
		\endgroup
    \caption{Mean brush centre distance from the grafting point perpendicular to the interface, $\braket{z}$, and along the shear flow, $\braket{x}$. The components of the inertia tensor are also shown.}
    \label{Rg}

\end{suppfigure}



\bibliography{manuscript}

\end{document}